# Scalable implementations of mean-field and correlation methods based on Lie-algebraic similarity transformation of spin Hamiltonians in the Jordan–Wigner representation

Shadan Ghassemi Tabrizi,[1,2*] Thomas M. Henderson,[3,4] Thomas D. Kühne,[2] and Gustavo E. Scuseria[3,4]

[1]*Computational System Sciences, Technische Universität Dresden, 01187 Dresden, Germany*

[2]*Center for Advanced Systems Understanding (CASUS), Am Untermarkt 20, 02826 Görlitz, Germany, *s.ghassemi-tabrizi@hzdr.de*

[3]*Department of Physics and Astronomy, Rice University, Houston, TX 77005-1892, USA*

[4]*Department of Chemistry, Rice University, Houston, TX 77005-1892, USA*

**Abstract.** Recent work has highlighted that the strong correlation inherent in spin Hamiltonians can be effectively reduced by mapping spins to fermions via the Jordan–Wigner transformation (JW). The Hartree–Fock method is straightforward in the fermionic domain and may provide a reasonable approximation to the ground state. Correlation with respect to the fermionic mean-field can be recovered based on Lie-algebraic similarity transformation (LAST) with two-body correlators. Specifically, a unitary LAST variant eliminates the dependence on site ordering, while a non-unitary LAST yields size-extensive correlation energies. Whereas the first recent demonstration of such methods was restricted to small spin systems, we present efficient implementations using analytical gradients for the optimization with respect to the mean-field reference and the LAST parameters, thereby enabling the treatment of larger clusters, including systems with local spins $s > \frac{1}{2}$.

## 1. Introduction

The physics of exchange-coupled systems is frequently described in terms of interacting spin centers.[1–3] As an example, the isotropic Heisenberg model, $H = \sum_{m<n} J_{mn} \mathbf{s}_m \cdot \mathbf{s}_n$, containing pairwise scalar couplings between site spins $\mathbf{s}_m = (s_m^x, s_m^y, s_m^z)$, can be used to represent the properties of multinuclear transition-metal complexes or magnetic solids. Because the state-space dimension grows rapidly with the number of sites, approximations are required[4–6] to obtain low-lying states (for interpreting spectroscopic data) or to compute thermal averages (e.g., temperature-dependent magnetic susceptibilities). Mean-field techniques belong to the less computationally demanding end of the spectrum of methods. Among these is projected Hartree–Fock theory[7] (PHF), where a three-dimensional spin configuration[8] that breaks spin and point-group symmetry is optimized with respect to projectors that restore these symmetries, thereby affording a state with definite quantum numbers. The loss of accuracy with increasing system size (PHF is not size-extensive) can be prevented by constructing a cluster basis[9,10] or by including additional configurations in the broken-symmetry reference.[11] Note that several other correlated methods widely used in quantum chemistry, such as many-body perturbation theory, coupled-cluster theory or configuration interaction, can be applied to spin systems as well,[12,13] preferably based on a cluster mean-field reference.[14] Although these approaches can adopt certain technical features of quantum-chemical implementations, they still fundamentally work in the spin representation, where the simplest mean-field state (a direct product of local spins) provides a poor approximation to the ground state.

In contrast, mapping spins to fermions via the Jordan–Wigner transformation[15] (JW) reduces strong correlations at the mean-field level.[16] Consequently, significantly more correlation is captured compared to a HF approach in the spin picture. Applications are not limited to one-dimensional systems with nearest-neighbor coupling (JW strings vanish in chains and reduce to a particle-number-depending sign factor in rings[17]): on general connectivity graphs, e.g., two- or three-dimensional lattices, the nonlocal JW strings can be evaluated as simple Thouless rotations[18] of a Slater determinant. Therefore, no mean-field decoupling of many-body terms arising from string operators is required.[16] In any case, such a decoupling is only practically feasible in special cases where a formulation of strings (Eq. (4) below) as $\prod_{q<p}(1-i\pi n_q)$ does not generate three- or higher-body terms in the Hamiltonian, see, e.g., Refs. 19–21. Moreover, it was recently demonstrated that a site-order independent formulation can be derived by generalizing the transformation towards extended JW (EJW[22,23]) and treating the additional degrees of freedom afforded by EJW as variational parameters corresponds to a unitary Lie-algebraic similarity-transformation (uLAST) approach;[24] when combined with an optimization of the Slater determinant, this becomes orbital-optimized uLAST (oo-uLAST).[24] These features make oo-uLAST an attractive foundation for correlation methods like non-unitary LAST (in the following simply referred to as LAST).[24,25]

Here we present efficient formulations for gradient-based HF, oo-uLAST and subsequent LAST in JW-transformed spin Hamiltonians. This enables the treatment of larger clusters than previously feasible by these methods. In the following Theory section, we briefly revisit the standard JW-transformation and its extended variant. As noted, regarding the EJW phase angles as variational parameters – motivated by the fact that EJW can be derived from uLAST – eliminates the dependence on site ordering and can, in certain cases, when HF orbitals become complex, provide additional correlation.[24] We derive analytic gradients for oo-uLAST and the analytic Jacobian for the equations determining the LAST amplitudes. The EJW angles (the uLAST parameters) and the reference determinant (the occupied orbitals) are optimized simultaneously in oo-uLAST, and LAST is employed to obtain size-extensive correlation on top of an oo-uLAST solution. In the Results section, oo-uLAST and LAST are benchmarked against exact-diagonalization (ED) data for larger systems than previously studied in the frame of an initial implementation[24] which operated in the full configuration-interaction (FCI) basis. Local spins $s > \frac{1}{2}$ are treated by decomposing them into spin-1/2 degrees of freedom, which are then fermionized through EJW. Finally, we discuss how these methods can serve as a basis for recovering additional correlation.

## 2. Theory

The JW-transformation maps spins to fermions, or vice versa. As an example, fermionic Hamiltonians are expressed in terms of qubit operators for quantum-computing applications. Here, we use the reverse direction,

$$s_p^+ = c_p^\dagger \phi_p^\dagger , \quad (1)$$

$$s_p^- = c_p \phi_p , \quad (2)$$

$$s_p^z = n_p - \tfrac{1}{2} , \quad (3)$$

because the fermionic domain typically mitigates strong correlation compared to the spin representation, allowing a mean-field treatment based on a single Slater determinant to provide a qualitatively reasonable approximation of the ground state.[24] In Eqs. (1) and (2), $s_p^+ = s_p^x + is_p^y$ and $s_p^- = s_p^x - is_p^y$ are spin ladder operators at site $p$, and $c_p^\dagger$ ($c_p$) are fermion creation (annihilation) operators. Eq. (4) specifies the standard JW string,

$$\phi_p^\dagger = \prod_{q<p} e^{i\pi n_q} . \quad (4)$$

Because the total fermion number, $\sum_p n_p = \sum_p c_p^\dagger c_p$, corresponds (up to a constant) to the $z$-component of the total spin, $S_z = \sum_p s_p^z$, cf. Eq. (3), the HF approximation is limited to Hamiltonians with $S_z$ symmetry, like the *XXZ* or $J_1 - J_2$ models studied here. In the fully anisotropic case, where no spin component is conserved and thus number symmetry and number parity are broken, more general mean-field methods would be needed.[26,27]

EJW represents a generalization of Eq. (4), which still ensures the correct commutation (anticommutation) relations for spins (fermions) by introducing the real parameters $\theta_{pq}$ subject to the constraints $|\theta_{pq} - \theta_{qp}| = \pi$, $\theta_{pp} = 0$,[a] see Eq. (5),

$$\phi_p^\dagger = e^{i \sum_q \theta_{pq} n_q} . \tag{5}$$

Eq. (6) shows the transformation of the $x$- and $y$-coupling contributions for a pair $\langle m, n \rangle$.

$$s_m^x s_n^x + s_m^y s_n^y = \tfrac{1}{2}\left( s_m^+ s_n^- + s_n^+ s_m^- \right) = \tfrac{1}{2}\left( c_m^\dagger \phi_m^\dagger c_n \phi_n + c_n^\dagger \phi_n^\dagger c_m \phi_m \right) \tag{6}$$

To treat the strings as Thouless rotations[18] of a Slater determinant [see Eq. (A12) in the Appendix A1] they are shifted to the right according to Eq. (7),

$$\phi_m^\dagger c_n = e^{i \sum_q \theta_{mq} n_q} c_n = c_n e^{i \sum_{q \neq n} \theta_{mq} n_q} , \tag{7}$$

yielding Eq. (8),

$$s_m^x s_n^x + s_m^y s_n^y = \tfrac{1}{2}\left[ c_m^\dagger c_n \phi_{m(n)}^\dagger \phi_{n(m)} + \text{h.c.} \right] , \tag{8}$$

where $\phi_{m(n)}^\dagger \equiv e^{i \sum_{q \neq n} \theta_{mq} n_q}$. Eq. (9) formulates $z$-coupling in normal order.

$$s_m^z s_n^z = n_m n_n - \tfrac{1}{2}(n_m + n_n) + \tfrac{1}{4} = c_m^\dagger c_n^\dagger c_n c_m - \tfrac{1}{2}(n_m + n_n) + \tfrac{1}{4} \tag{9}$$

Thus, in either standard JW or EJW, the spin Hamiltonian[b] can be faithfully mapped onto the fermionic Hamiltonian of Eq. (10),

$$\sum_{m<n} J_{mn} \left\{ \tfrac{1}{2}\left[ c_m^\dagger c_n \phi_{m(n)}^\dagger \phi_{n(m)} + \text{h.c.} \right] + n_m n_n - \tfrac{1}{2}(n_m + n_n) + \tfrac{1}{4} \right\} . \tag{10}$$

---

[a] Without loss of generality, we may set $\theta_{qp} = \theta_{pq} + \pi$ for $p < q$. The EJW strings are unitary and commute with their associated creation and annihilation operators, e.g., $[c_p, \phi_p] = 0$. However, in contrast to standard JW, EJW strings are generally not Hermitian, i.e., $\phi_p^\dagger \neq \phi_p$.

[b] For the sake of a slightly simpler notation, we focus here on the Heisenberg model instead of the *XXZ* model.

Consistency with the convention of fully antisymmetrized two-particle integrals in a second-quantized Hamiltonian[28] is established through Eq. (11):

$$c_m^\dagger c_n^\dagger c_n c_m = \tfrac{1}{4}\left(c_m^\dagger c_n^\dagger c_n c_m + c_n^\dagger c_m^\dagger c_m c_n - c_m^\dagger c_n^\dagger c_m c_n - c_n^\dagger c_m^\dagger c_n c_m\right) . \tag{11}$$

This allows the $z$-coupling part, $H_z$, to be written in a standard form, Eq. (12), with appropriately chosen values for the one- and two-particle integrals, $t_{kl}$ and $[kn|lm]$, respectively, and $E_{\text{const}} = \tfrac{1}{4}\sum_{m<n} J_{mn}$. We adopt this formulation in the derivations provided in the Appendix.

$$H_z \equiv \sum_{m<n} J_{mn} s_m^z s_n^z = E_{\text{const}} + \sum_{kl} t_{kl} c_k^\dagger c_l + \frac{1}{2}\sum_{klmn}[kn|lm]c_k^\dagger c_l^\dagger c_m c_n \tag{12}$$

**HF and oo-uLAST.** The HF method consists in variationally optimizing a Slater determinant $|\Phi\rangle$. In quantum-chemical terminology, $a_o^\dagger$ acting on the vacuum $|0\rangle$ in Eq. (13) creates a fermion in a molecular orbital (indices $o$ and $v$ will be used for occupied and virtual MOs, respectively),

$$|\Phi\rangle = \left(\prod_{o \in \text{occ}} a_o^\dagger\right)|0\rangle . \tag{13}$$

However, for fixed strings (e.g., standard JW), the HF energy depends on the site numbering. This might appear to be problematic for non-linear lattices only, but even in *XXZ* rings or chains, depending on the $\Delta$ parameter (see Eq. (29) below), a successive ordering may not be optimal in terms of the HF energy.[24] Treating EJW-angles $\theta_{pq}$ as optimization parameters corresponds to a special form of uLAST and makes the ansatz independent of the site numbering.[24] The simplest unitary or non-unitary LAST variants employ a two-body correlator,[25] Eq. (14):

$$\gamma_2 = \tfrac{1}{2}\sum_{p,q}\gamma_{pq} n_p n_q . \tag{14}$$

Only the $\gamma_{pq}$ parameters with $p < q$ are independent. When setting $\gamma_{pq} = i\theta_{pq}$, with real $\theta_{pq}$, $e^{\gamma_2}$ becomes unitary. By imposing the constraints $|\theta_{pq} - \theta_{qp}| = \pi$ and $\theta_{pp} = 0$, the EJW-transformed Hamiltonian is obtained from standard JW through Eq. (15),

$$H_{\text{EJW}} = e^{-\gamma_2} H_{\text{JW}} e^{\gamma_2} . \tag{15}$$

Henceforth, it will be clear from the context whether $H$ represents $H_{\text{EJW}}$ or $H_{\text{JW}}$ (an explicit differentiation is usually not necessary). In the following two sections and in the Appendix (A1 and A2), we discuss the simultaneous optimization (using analytical gradients) of the Slater determinant and the angles $\theta_{pq}$, which constitutes the oo-uLAST method.

For a given $H$, the optimization of $|\Phi\rangle$ corresponds to a HF procedure. $|\Phi\rangle$ is defined in terms of a Thouless rotation acting on an initial guess $|\Phi^0\rangle \equiv \prod_o (a_o^0)^\dagger |0\rangle$, Eq. (16),

$$|\Phi\rangle = N(Z)e^Z |\Phi^0\rangle , \qquad (16)$$

where $N(Z)$ is a normalization constant, and $Z$ is defined in Eq. (17):

$$Z = \sum_{v \in \text{virt}} \sum_{o \in \text{occ}} Z_{vo} a_v^\dagger a_o . \qquad (17)$$

The minimization of the HF energy, Eq. (18),

$$E = \frac{\langle \Phi^0 | e^{Z^\dagger} H e^Z | \Phi^0 \rangle}{\langle \Phi^0 | e^{Z^\dagger} e^Z | \Phi^0 \rangle} , \qquad (18)$$

with respect to the amplitudes $Z_{vo}$ is detailed in Appendix A1, and the uLAST optimization of the angles $\theta_{pq}$ is explained in Appendix A2.

**LAST.** As in Ref. 24, we apply non-unitary LAST to capture additional correlation energy for an oo-uLAST solution. We derive and implement the analytic Jacobian for the LAST amplitudes, enabling an efficient optimization procedure. The two-body correlator is parametrized by a real symmetric matrix $\boldsymbol{\alpha}$, with $\alpha_{pp} = 0$, Eq. (19),

$$\alpha_2 = -\tfrac{1}{2} \sum_{p,q} \alpha_{pq} n_p n_q , \qquad (19)$$

The respective transformations of creation and annihilation operators are given in Eqs. (20) and (21).[24,25] Note that $\bar{c}_p^\dagger$ and $\bar{c}_p$ are not Hermitian conjugates of each other.

$$\bar{c}_p^\dagger \equiv e^{-\gamma_2} c_p^\dagger e^{\gamma_2} = c_p^\dagger e^{\sum_q \gamma_{pq} n_q} \qquad (20)$$

$$\bar{c}_p \equiv e^{-\gamma_2} c_p e^{\gamma_2} = c_p e^{-\sum_q \gamma_{pq} n_q} \qquad (21)$$

The transformed Hamiltonian of Eq. (22),

$$\bar{H} = e^{-\alpha_2} H e^{\alpha_2} , \qquad (22)$$

is still given by Eq. (10) above, but now the strings take the form of Eq. (23):

$$\phi_p^\dagger = e^{\sum_p (\alpha_{pq} + i\theta_{pq}) n_q} . \tag{23}$$

Using LAST on top of oo-uLAST means that the mean-field state $|\Phi\rangle$ and uLAST parameters $\boldsymbol{\theta}$ remain fixed, and $\boldsymbol{\alpha}$ is optimized to solve the system of equations (24) for the residuals $R_{pq}$ of all independent index pairs $(p < q)$,[25]

$$R_{pq} \equiv \langle \Phi | n_p n_q (\bar{H} - E) | \Phi \rangle = 0 , \tag{24}$$

where the energy is given in Eq. (25):

$$E = \frac{\langle \Phi | \bar{H} | \Phi \rangle}{\langle \Phi | \Phi \rangle} . \tag{25}$$

It is useful to pass the Jacobian, defined in Eq. (26), to the numeric solver (in our `Matlab` program, we use the `fsolve` function).

$$J_{pq}^{rs} \equiv \frac{\partial R_{pq}}{\partial \alpha_{rs}} \tag{26}$$

Equations for the computation of the energy, the residuals and the Jacobian are provided in Appendix A3.

**Spin-1/2 decomposition for $s > 1$.** A well-known example of representing on-site spins $s > \tfrac{1}{2}$ by auxiliary spin-1/2 degrees of freedom (qubits) is the Affleck–Kennedy–Lieb–Tasaki[29] (AKLT) construction for the $s = 1$ chain: each physical spin is modeled by two qubits, singlet bonds are formed between qubits on neighboring sites, and sites are finally projected onto their local triplet subspace. Such constructions have become a conceptual tool in the theory of valence-bond solids. As examples of applications of spin-1/2 decompositions beyond such settings, we mention a technique to constrain the PHF mean-field reference to represent a spin-coherent product state,[8] and the selection of spin configurations that form a linearly independent and complete set of spin eigenfunctions upon projection onto a total-spin subspace.[30] Here, we introduce spin-1/2 particles, Eq. (27),

$$\mathbf{s}_p \to \sum_{a=1}^{2s_p} \boldsymbol{\kappa}_{p,a} , \tag{27}$$

to bring systems with local spins $s_p > \tfrac{1}{2}$ into the fermionic mean-field framework by applying a JW- or EJW-transformation to the qubits $\boldsymbol{\kappa}_{p,a}$. We are aware of only a few previous works where a similar decomposition combined with a JW mapping has been used, e.g., anisotropy-dependent phase diagrams and the Haldane gap in $s=1$ chains were studied in Refs. 31,32. To our knowledge, there are no previous applications of this strategy for $s>1$. Even for an $s=1$ chain, the JW mapping already carries nonlocal strings (in contrast to the $s=\frac{1}{2}$ chain). Then, the coupling between most distant qubit pairs, i.e., the first and the second qubits of sites $m$ and $m+1$, $\kappa_{m,1}$ and $\kappa_{m+1,2}$, respectively (see Eq. (28) below), generates six-body terms when expanding string factors as $e^{i\pi n_p} = 1 - i\pi n_p$.[c] Therefore, unless one evaluates the strings exactly as Thouless rotations,[16] as we do here, a mean-field decoupling would generally become unwieldy for $s > \frac{1}{2}$, even in one-dimensional systems.

In the enlarged Hilbert space of $\tilde{H}$, Eq. (28),

$$H = \sum_{m<n} J_{mn} \mathbf{s}_m \cdot \mathbf{s}_n \rightarrow \tilde{H} = \sum_{m<n} J_{mn} \sum_{a,b} \boldsymbol{\kappa}_{m,a} \cdot \boldsymbol{\kappa}_{n,b} , \qquad (28)$$

the $\kappa_{p,a}$ can be coupled into on-site spin values that are smaller than $s_p$. Enforcing the physical sector of maximal local spins would require full symmetrization of the auxiliaries at every site, which is impractical. Besides the rapidly increasing length of the respective linear combinations with the number of sites, permutations of spin-1/2 particles generally do not correspond to simple orbital permutations in the JW picture (this contrasts with the spin representation used in PHF,[8] where symmetry projection is achieved by Thouless rotations of orbitals). However, contamination by unphysical states with lower on-site spins does not affect the variational principle for the ground state, because the latter has maximal local spins, as proven in Appendix A4. Consequently, the ground state – and, indeed, the ground state in each magnetization sector of the *XXZ* model – is fully embedded in the correct state space, i.e., any trial wave function in $\tilde{H}$ is variationally consistent with $H$.

Although not considered here, we would like to mention another fermionization scheme for $s > \frac{1}{2}$ proposed by Batista and Ortiz.[33] It associates the $2s+1$ states of each site with the occupation state of an orbital, which may be empty or host a fermion with $2s$ different flavors. However, in contrast to the auxiliary spin-1/2 decomposition, unphysical contributions (corresponding to multiply occupied sites) may affect the validity of the variational principle for the ground state and would have to be excluded.

---

[c] In a conventional successive numbering of spin-1/2 particles, the *x,y*-coupling between $\kappa_{m,1}$ and $\kappa_{m+1,2}$ involves a string $e^{i\pi(n_{m,2}+n_{m+1,1})}$. The *x,y*-coupling thus contributes $c_{m,1}^\dagger c_{m+1,2} n_{m,2} n_{m+1,1}$, among other terms.

Lastly, the constraint-free mapping for spins with multiplicities $(2s+1) = 2^n$ suggested by Dobrov[34] is based on the equivalence of the dimensions of spin and fermion spaces. While it reduces to standard JW for $s = \frac{1}{2}$, several difficulties occur for $s > \frac{1}{2}$. In this scheme, even the isotropic Heisenberg model breaks fermion-number parity, which would require specialized mean-field approaches[26,27] involving Hartree–Fock–Bogoliubov[35] (HFB) states, where matrix elements between HFB states may have to be computed in a numerically robust formulation.[36] Another hurdle for practical applications are high-rank fermionic terms, e.g., pairwise couplings between $s = \frac{7}{2}$ sites afford five-body interactions (not counting JW strings).

## 3. Results

To explore systems beyond the size accessible to an initial testing implementation,[24] we consider isotropic rings with up to 30 sites. We also explore the performance of JW-HF, oo-uLAST and LAST for the *XXZ* model,

$$H = \sum_{\langle m,n \rangle} \left( s_m^x s_n^x + s_m^y s_n^y + \Delta s_m^z s_n^z \right), \quad (29)$$

where the sum in Eq. (29) runs over all pairs $\langle m,n \rangle$ of nearest neighbors, as well as for the isotropic $J_1 - J_2$ model, Eq. (30),

$$H = J_1 \sum_{\langle m,n \rangle} \mathbf{s}_m \cdot \mathbf{s}_n + J_2 \sum_{\langle\langle m,n \rangle\rangle} \mathbf{s}_m \cdot \mathbf{s}_n, \quad (30)$$

which includes both antiferromagnetic NN $(J_1 > 0)$ and next-nearest-neighbor (NNN) couplings $(J_2 > 0)$. We cover local spins up to $s = 3$ in selected cases. Rings, chains, and square lattices with open (OBC) or periodic (PBC) boundary conditions are chosen so that their ground-state energies are available from (sparse-matrix) ED techniques or Bethe-ansatz calculations, or from results reported in the literature. In addition, for rings and an icosidodecahedron, we compare field-dependent magnetization curves (which are determined by energy differences between different fermion-number sectors) against exact results.

**Isotropic rings.** To demonstrate that size-consistent variational oo-uLAST outperforms correlated variational approaches that lack size consistency for large enough systems, we specifically compare to earlier PHF results for Heisenberg spin rings.[8] Table 1 compares oo-uLAST energies with reference values from ED or density-matrix renormalization group (DMRG) calculations (taken from Refs. 37 or 38) for the $S = 0$ ground states of antiferromagnetic $(J = 1)$ rings with on-site spins $s = \frac{1}{2}, 1, \frac{3}{2}, 2, \frac{5}{2}$ and ring sizes $N = 6, 12, 18, 24, 30$. The largest system, $N = 30$, $s = \frac{5}{2}$, comprises 150 auxiliary spin-1/2 particles, significantly exceeding the ~16 such particles that could previously be treated within a full CI basis.[24]

Table 1: Ground-state energies of antiferromagnetic Heisenberg rings from oo-uLAST and ED (or DMRG).[a]

| $s$ | $N$ | | | | | |
| --- | --- | --- | --- | --- | --- | --- |
| | 6 | 12 | 18 | 24 | 30 | method |
| 1/2 | -2.803 | -5.387 | -8.023 | -10.670 | -13.322 | exact |
| | -2.667 | -5.193 | -7.782 | -10.375 | -12.969 | oo-uLAST |
| 1 | -8.617 | -16.870 | -25.242 | -33.641 | -42.046 | exact/DMRG |
| | -7.959 | -15.886 | -23.904 | -31.953 | -39.993 | oo-uLAST |
| 3/2 | -17.393 | -34.131 | -51.031 | -67.968 | -84.919 | exact/DMRG |
| | -16.163 | -32.352 | -48.651 | -64.976 | -81.335 | oo-uLAST |
| 2 | -29.165 | -57.408 | -85.873 | -114.390 | -142.927 | exact/DMRG |
| | -27.387 | -54.817 | -82.383 | -109.979 | -137.586 | oo-uLAST |
| 5/2 | -43.935 | -86.679 | -129.703 | -172.793 | -215.909 | exact/DMRG |
| | -41.612 | -83.286 | -125.117 | -166.962 | -208.667 | oo-uLAST |

[a] For systems that were too large for ED, DMRG energies were taken from Ref. 38. All ED/DMRG entries agree with Table 5.1 in the latter work.

The fractional correlation energy $p$ is defined as the ratio of Eq. (31),

$$p = \frac{E_{\text{method}} - E_{\text{classical}}}{E_{\text{exact}} - E_{\text{classical}}} \, , \qquad (31)$$

where $E_{\text{classical}}$ corresponds to the energy of the best classical spin configuration; for rings with even N, $E_{\text{classical}} = -Ns^2$ (Néel configuration).

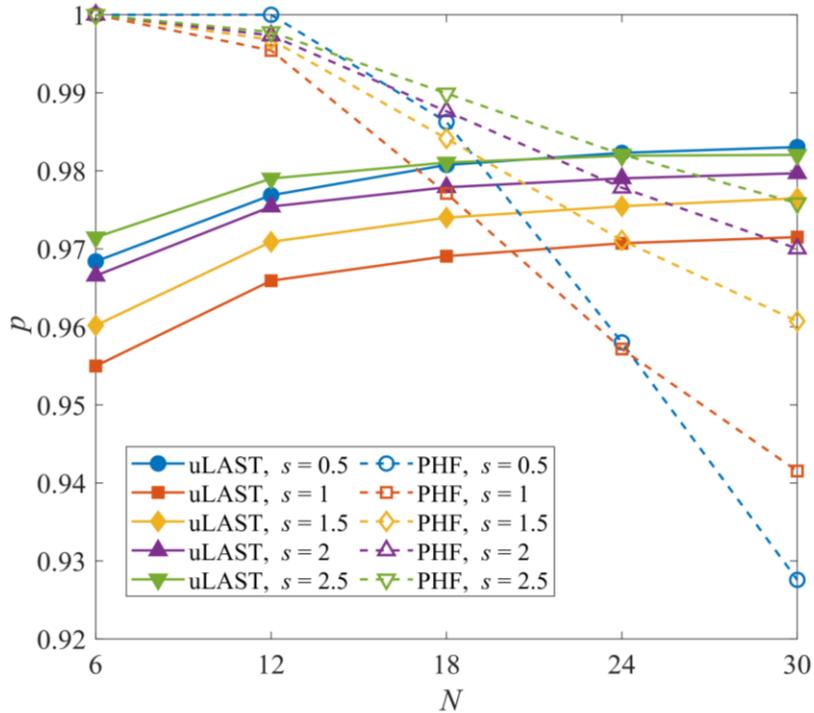

Figure 1: Fractional correlation energy $p$ from oo-uLAST or PHF for Heisenberg rings. See text for details.

Figure 1 compares oo-uLAST to PHF (employing spin and point-group symmetry, data taken from Ref. 8). For each $s$, the oo-uLAST curves increase smoothly with $N$ and appear to converge around $p = 0.97$–$0.99$, although definitive conclusions would require consideration of larger rings. These observations highlight the inherent size extensivity of oo-uLAST. By contrast, the PHF results are exact or nearly exact for $N = 6$ for all $s$ values but decline monotonically as $N$ increases, manifesting the lack of size extensivity.[7,39] The energy gained from symmetry restoration scales sub-linearly with system size, resulting in a progressive dilution of the recovered correlation energy as additional sites are introduced. This effect is most pronounced for $s = \frac{1}{2}$ (compare $N = 6$ to $N = 30$) but remains evident in the more classical regime of larger $s$ values; for $s = \frac{5}{2}$, oo-uLAST surpasses PHF around $N = 24$.

*XXZ* **model.** We consider several one- and two-dimensional systems with different on-site spins $s$. To obtain the optimal oo-uLAST solution across the range $-2 \leq \Delta \leq 2$, we first independently converge solutions, using random initial guesses, for selected $\Delta$ values. From each of these points, we proceed stepwise in both directions, always using the solution from the neighboring $\Delta$ value as an initial guess. For each $\Delta$, the solution with the lowest energy serves as the reference state for a similarity-transformed LAST calculation (the same procedure is applied for the parameter scans in the $J_1 - J_2$ model, see the following section).

The JW representation reduces the strong correlation characteristic of the spin domain, so that even a single Slater determinant may serve as a reasonable approximation to the ground state. This was demonstrated in Ref. 16, where the overlap between the exact ground state and the JW-HF wavefunction was computed in *XXZ* chains. In Figure 2, we additionally examine the respective overlap for oo-uLAST solutions in $N = 6$ and $N = 12$ spin-1/2 chains. The maximum configuration-interaction (CI) coefficient is defined as the absolute value of the overlap between a mean-field wave function (JW-HF or oo-uLAST) and the exact ground state. In the spin representation, this coefficient is taken to represent the largest possible overlap between any collinear spin configuration and the ground state. We do not illustrate the dependence of the JW-HF energy on the numbering scheme for spin-1/2 sites,[24] and instead compare oo-uLAST only to JW-HF based on a sequential numbering along the chain. A maximal CI coefficient close to 1 indicates that a single determinant represents a qualitative reasonable approximation. For $\Delta < 0$, oo-uLAST is distinct from JW-HF (thus yielding better energies), and its overlap with the exact solution is larger. In the limits of large positive or negative $\Delta$, the ground state becomes two-fold degenerate (the largest CI coefficient therefore approaches $1/\sqrt{2}$), which we resolve in the ED routine by adapting the Hamiltonian to spin-flip symmetry.

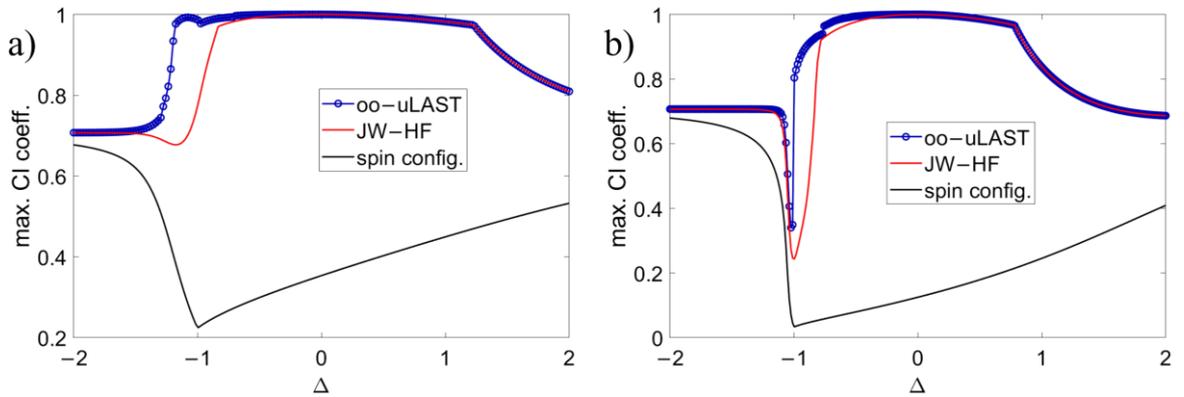

Figure 2: Maximum CI coefficients as functions of $\Delta$ for an *XXZ* chains with $N = 6$ (panel a) or $N = 12$ (panel b) spin-1/2 sites. See text for details.

Progressions of local spin values, $s = \frac{1}{2}, 1, \frac{3}{2}, 2, \frac{5}{2}, 3$, are studied for a chain and a ring with $N = 4$ centers in Figure 3 and Figure 4, respectively, which show relative energy errors, $(E_{\text{method}} - E_{\text{exact}})/E_{\text{exact}}$, in the $M = 0$ sector as a function of $\Delta$. Note that, from a physical perspective, the *XXZ* model is not necessarily the most appropriate or complete anisotropic spin model for $s \geq 1$ systems. In such cases, axial zero-field splitting terms, $D_m(s_m^z)^2$, could be incorporated straightforwardly into the mean-field framework, but are omitted here for simplicity.

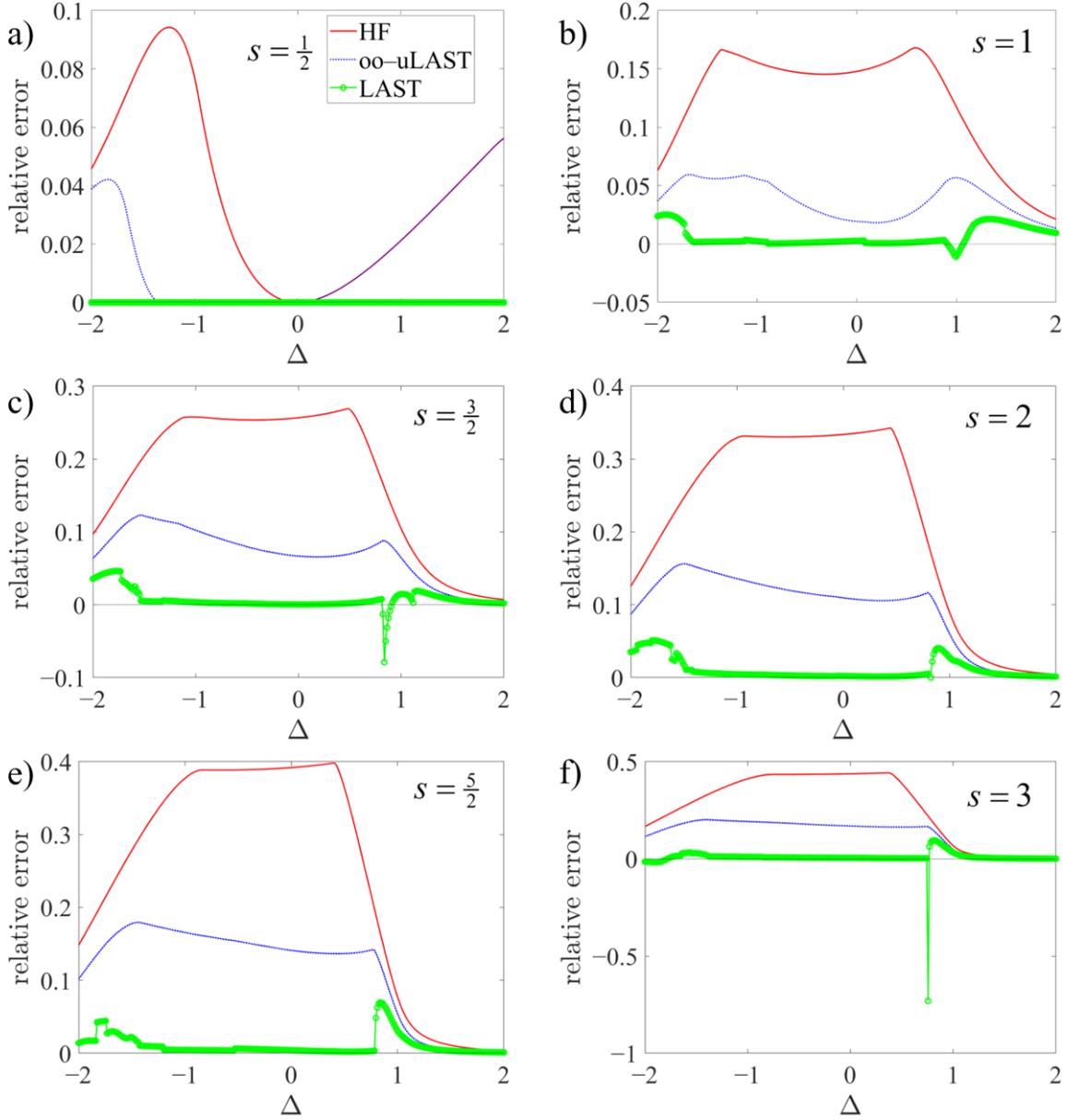

Figure 3: Relative energy errors for an $N = 4$ chain as a function of $\Delta$ in the *XXZ* model for with different on-site spin values *s*.

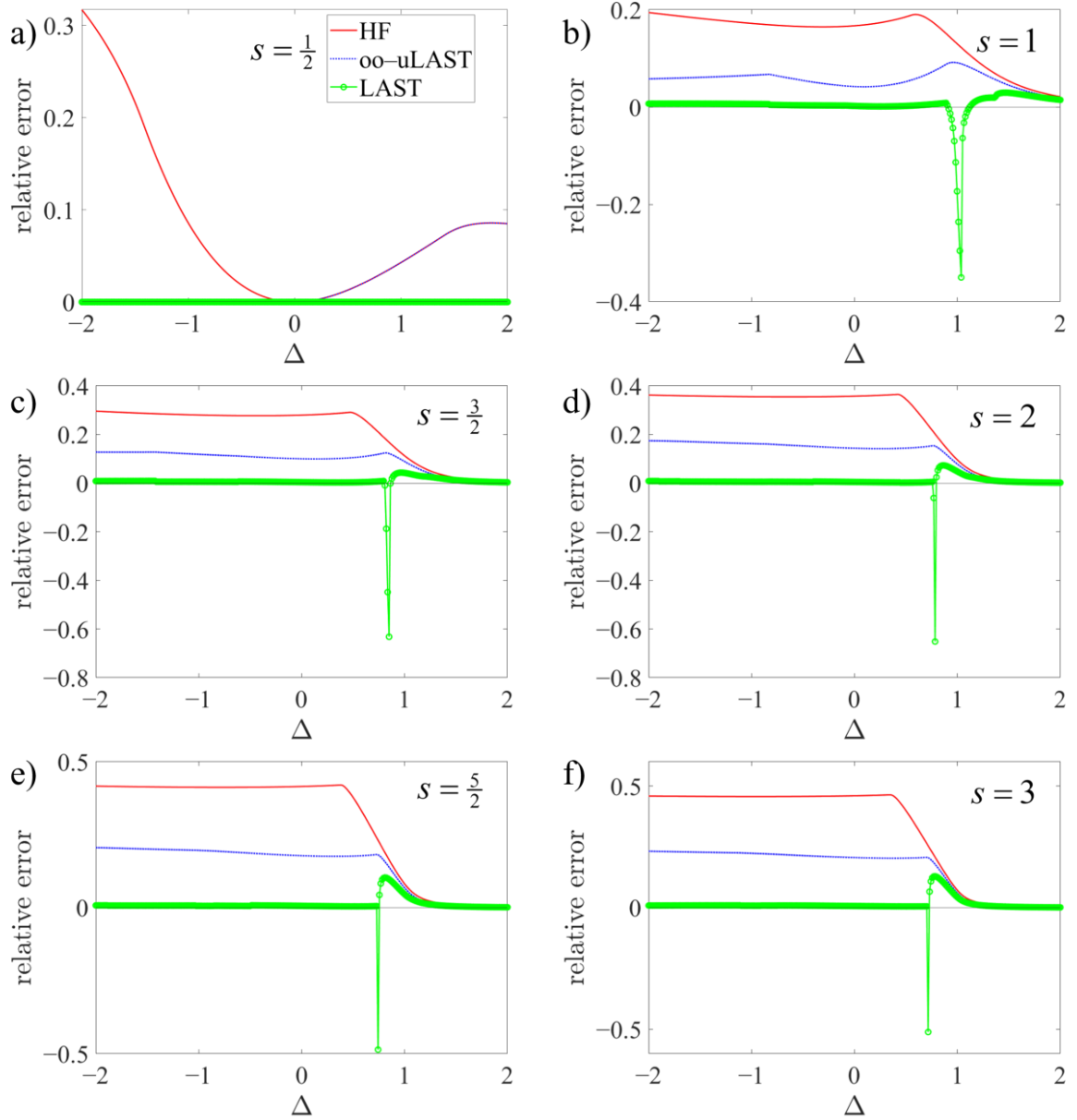

Figure 4: Relative energy errors for an $N = 4$ ring as a function of $\Delta$ in the *XXZ* model with different on-site spin values *s*.

We also illustrate JW-HF energies in Figure 3 and Figure 4. As shown in Figure 5, we first number the auxiliary spin-1/2 particles within a given center before proceeding to the next center. In general, however, this numbering (which is inconsequential for oo-uLAST) of the (auxiliary) spin-1/2 particles cannot be expected to be optimal for HF calculations within the standard JW transformation. In fact, we observe an equivalence between oo-uLAST and JW-HF only for the spin-1/2 ring and chain for $\Delta \geq 0$. Strictly speaking, this does not rule out the possibility that, in other systems or for different values of $\Delta$, oo-uLAST may still become equivalent to JW-HF with an alternative numbering of the spin-1/2 particles, a point we have not investigated here. For $N$ sites, there are $(2sN)!$ possible orderings of the auxiliaries. Despite intra-site permutations and permutations from lattice symmetries (mirror reflection in open chains, dihedral group $D_N$ in rings, etc.) being inconsequential, the number of inequivalent permutations is too large to examine individually in all but the smallest systems.

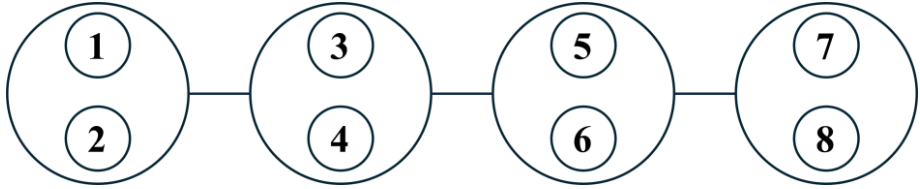

Figure 5: As a basis for a standard JW-transformation, we assign consecutive numbers to the auxiliary spins associated with $s=\frac{1}{2}$ sites (illustrated here for a chain with four $s=1$ sites).

In many instances, the LAST energy curves as a function of $\Delta$ exhibit numerous discontinuities and kinks, whereas JW-HF and oo-uLAST show virtually no discontinuities and significantly fewer kinks (the latter more so for oo-uLAST than for JW-HF) at points where different mean-field solutions cross, an observation consistent with earlier findings.[24] Since the resolution of the parameter scans is necessarily limited, we cannot make a definite statement as to whether jumps in the LAST energy curves correspond to genuine discontinuities or merely to sharp transitions. These irregularities arise from our present protocol, in which the energetically lowest oo-uLAST solution found for a given $\Delta$ is taken as the reference and LAST amplitudes $\boldsymbol{\alpha}$ are initialized at zero. Thus, convergence to a local rather than a global solution cannot be excluded. Starting from different $\boldsymbol{\alpha}$ guesses or taking an alternative, energetically nearby oo-uLAST reference could in principle lead to a better approximation to the energy and may yield a smoother curve. Indeed, it had already been noted in Ref. 24 that several oo-uLAST solutions often coexist, and there is no guarantee that the lowest of them provides the optimal reference for LAST. A more exhaustive strategy of optimizing all parameters simultaneously (the uLAST and LAST parameters, $\boldsymbol{\theta}$ and $\boldsymbol{\alpha}$, respectively, and the Slater determinant $|\Phi\rangle$) is however beyond the scope of the present work.

In any case, LAST typically recovers a substantial fraction of the remaining correlation energy. However, being non-variational, it occasionally collapses to spurious solutions below the exact ground state for $s \geq 1$ systems near the isotropic point $\Delta = 1$, most notably for the ring in Figure 4. The problematic region, however, becomes progressively narrower with increasing $s$: from $s=1$ to $s=3$, its extent shrinks such that, at our scan resolution, only a single point is affected for $s=\frac{5}{2}$ and $s=3$. Apart from this issue, LAST is virtually exact across the entire range for $s=\frac{1}{2}$, and for larger $s$ it remains highly accurate over a broad window around $\Delta=0$, where, particularly for higher spin values, the region following the irregularity near $\Delta=1$ is again described accurately at larger $\Delta$. Note that for $s>\frac{1}{2}$, the JW (or EJW) mapping introduces non-vanishing string operators even for a chain. As a result, at $\Delta=0$ the *XXZ* Hamiltonian no longer reduces to an effective non-interacting problem and thus admits no exact mean-field solution, in contrast to $s=\frac{1}{2}$ chains.

Figure 6 presents results for systems with 24 $s=\frac{1}{2}$ sites: a chain, a ring, and increasingly compact square lattices, namely $12\times 2$, $8\times 3$ and $6\times 4$, each studied with both OBC and PBC. The JW-HF results in $n_x \times n_y$ systems are based on a numbering according to Eq. (32), same as Eq. 37 in Ref. 24,

$$i_{x,y} = x + (y-1)n_x \qquad (32)$$

where $n_x > n_y$. The chain and the ring are non-interacting at $\Delta=0$, such that JW-HF with sequential site numbering (which eliminates the strings) is exact at this point, as are oo-uLAST and LAST. In addition, oo-uLAST appears to be exact also for the $12\times 2$ systems at $\Delta=0$. Overall, two-dimensional lattices are described less accurately than one-dimensional systems, and relative errors tend to be larger for PBC than for OBC. For instance, the relative error of oo-uLAST remains below 3% for the chain, whereas it exceeds 10% for the $6\times 4$ PBC lattice near $\Delta=-1$.

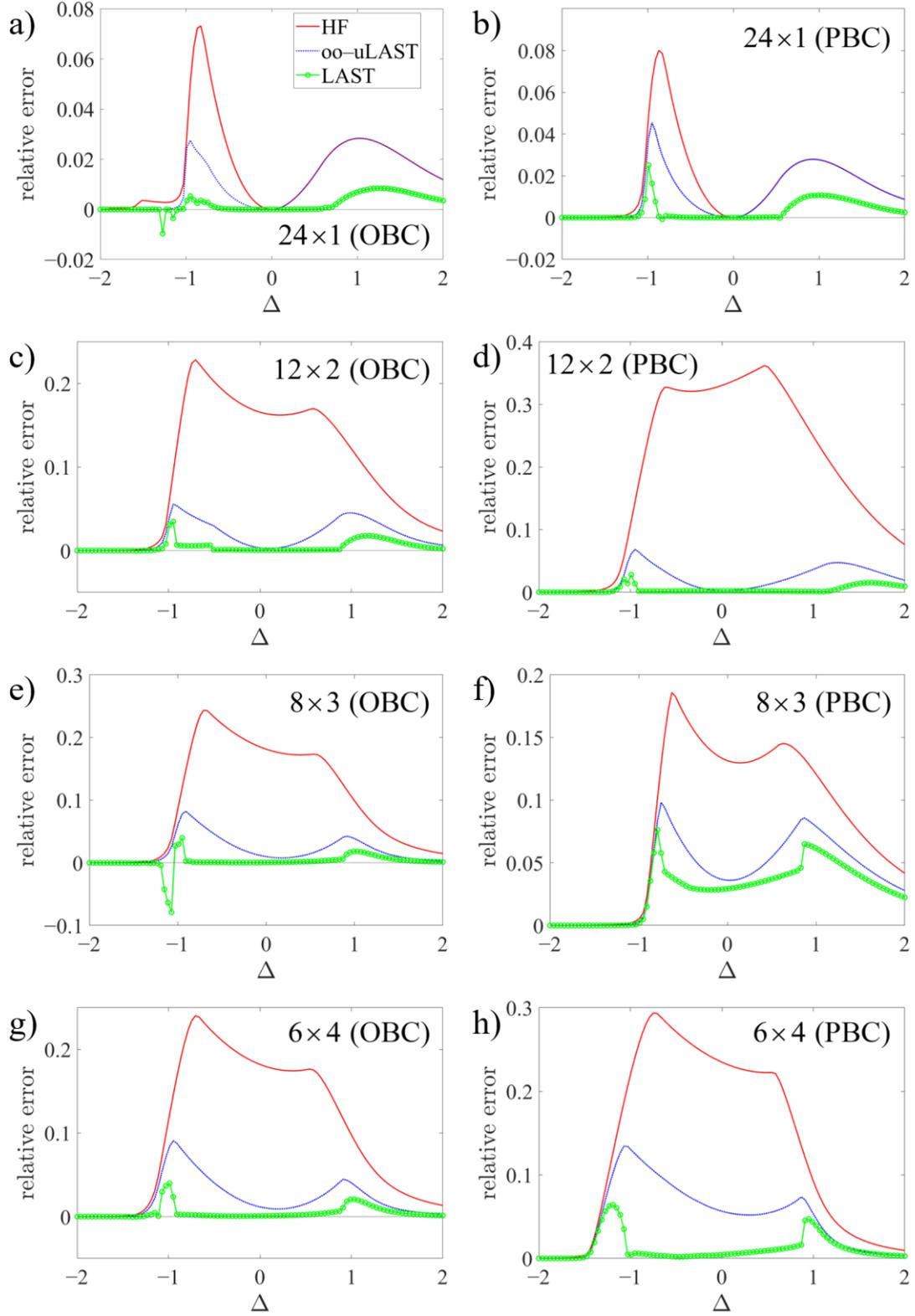

Figure 6: Relative energy errors for the *XXZ* model with 24 $s = \frac{1}{2}$ sites as a function of $\Delta$.

***J*₁–*J*₂ model.** For the same lattices with 24 $s = \tfrac{1}{2}$ sites that were used for the *XXZ* model in the previous section, results for the relative energy errors in the $J_1 - J_2$ model are plotted in Figure 7. For the chain, the LAST curve is notably jittery, reflecting multiple competing solutions; we have attempted to follow the lowest-energy branch by seeding each point with the converged solution at the neighboring value of $J_2 / J_1$, as described above for oo-uLAST. Despite this piecewise behavior, LAST is very accurate in a neighborhood of the Majumdar–Ghosh point $J_2 / J_1 = 0.5$,[40] where JW-HF is exact (in the chain and the ring) due to dimer formation. A striking positive case is the $12 \times 2$ PBC lattice, where LAST is essentially exact over a wide interval, extending up to $J_2 / J_1 \approx 0.85$. By contrast, on other lattices LAST exhibits more irregular features (cusps and small discontinuities), with a pronounced dip for $6 \times 4$ PBC around $J_2 / J_1 \approx 0.65$ that crosses slightly below zero error (non-variational behavior).

Overall, oo-uLAST provides far smoother curves, consistently better than JW-HF (again except for the chain and the ring, where oo-uLAST is equivalent to JW-HF for $J_2 / J_1 \leq 0.5$), and less erratic than LAST. As observed on the respective lattices in the *XXZ* model, two-dimensional lattices or PBC typically yield larger errors, with the $12 \times 2$ PBC exception highlighted above.

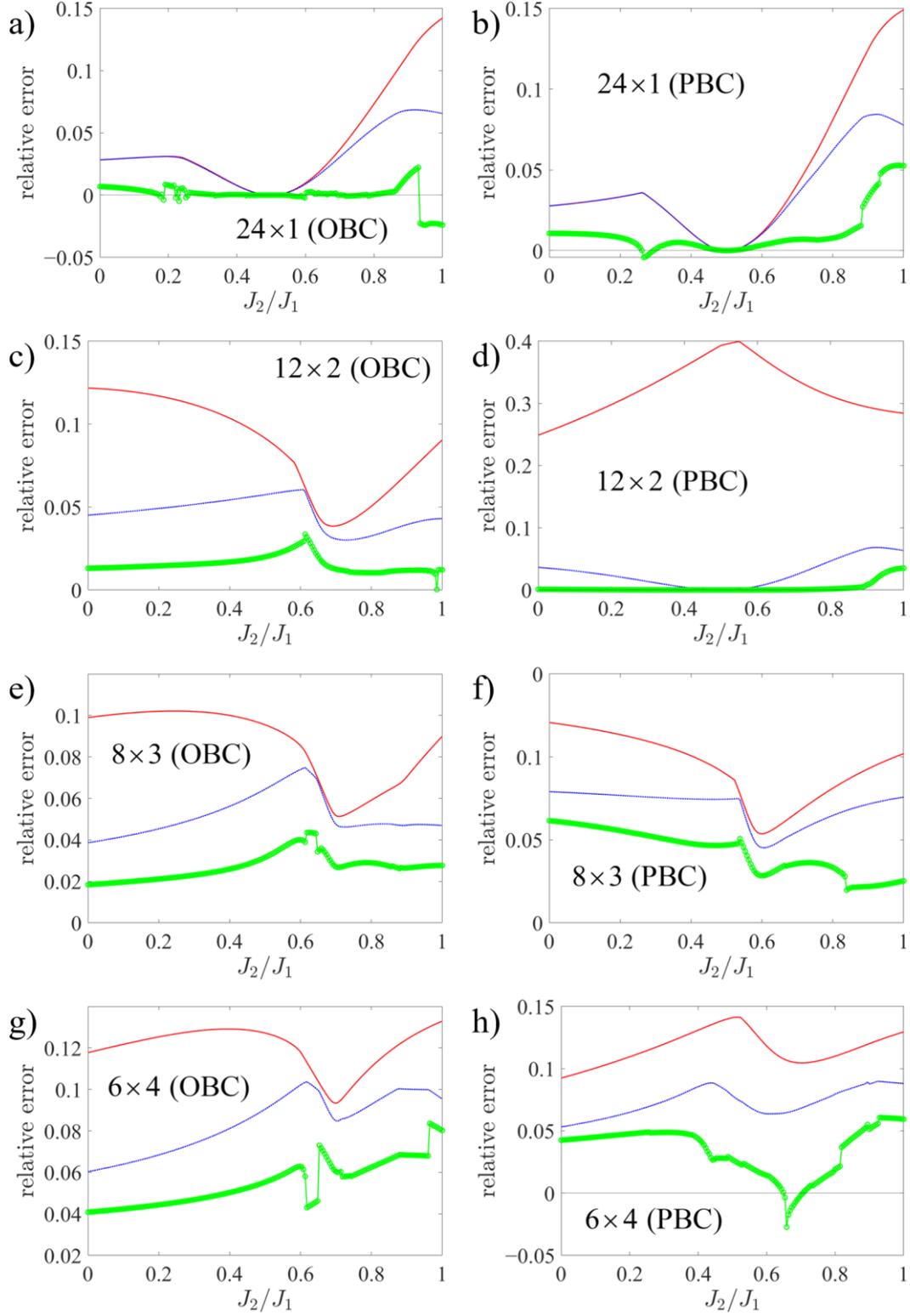

Figure 7: Relative energy errors for systems with 24 $s = \frac{1}{2}$ sites as a function of $J_2/J_1$ in the antiferromagnetic ($J_1 = 1$) $J_1 - J_2$ model. The lattices are the same as in Figure 6.

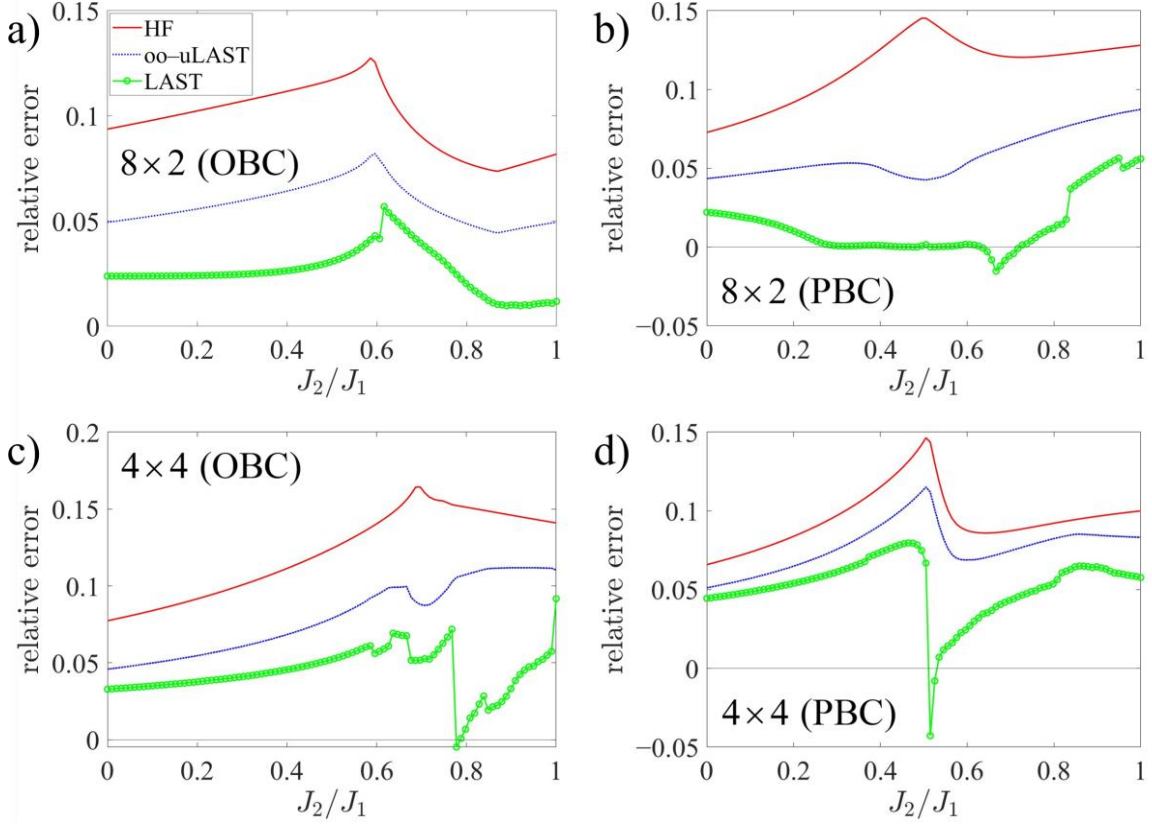

Figure 8: Relative energy errors for systems with 16 $s=1$ sites in the antiferromagnetic $J_1-J_2$ model.

Finally, results for $8\times 2$ and $4\times 4$ lattices composed of $s=1$ centers are shown in Figure 8, both with OBC or PBC. For $8\times 2$ with OBC, all three curves run approximately in parallel. In general, however, the LAST curves display markedly larger irregularities than for comparable spin-1/2 systems, which is especially true for the $4\times 4$ OBC case. The best description is obtained for $8\times 2$ with PBC, where, within an interval around $J_2/J_1 \approx 0.45$, LAST is nearly exact.

**Field-dependent magnetization.** So far, we have worked within the $M=0$ sector, which corresponds to half-filling in the JW representation. Here, we consider the field-dependent magnetization for antiferromagnetic isotropic $s=\tfrac{1}{2}$ rings and the icosidodecahedron, which is determined by energy differences between $M$ sectors (fermion-number sectors). At zero temperature, the thermal average is $\langle \mathcal{M} \rangle = gM$, where $g$ is the gyromagnetic factor and $M$ is the $S_z$ eigenvalue of the field-dependent global ground state. The levels $E_i = E_i^{(0)} - g\mu_B B M_i$ comprise a field-independent term $E_i^{(0)}$ and a Zeeman contribution (we set $g=2$ and $\mu_B = 1$).

Figure 9 and Figure 10 display $\langle \mathcal{M} \rangle$ as a function of the magnetic field $B$ for $N=30$ and $N=60$ rings, respectively. The exact curves were obtained from Bethe-ansatz energies computed using the ABACUS[41] program. In both systems, oo-uLAST reproduces the staircase structure only qualitatively: the first step from $M=0$ to $M=1$, corresponding to the singlet-triplet gap, is shifted to a significantly higher field, which compresses the spacing between the first and second step (at $B \approx 0.15$ for $N=60$). Adding correlation via LAST removes this defect almost completely for $N=30$, and LAST predictions for $N=60$ are also rather accurate.

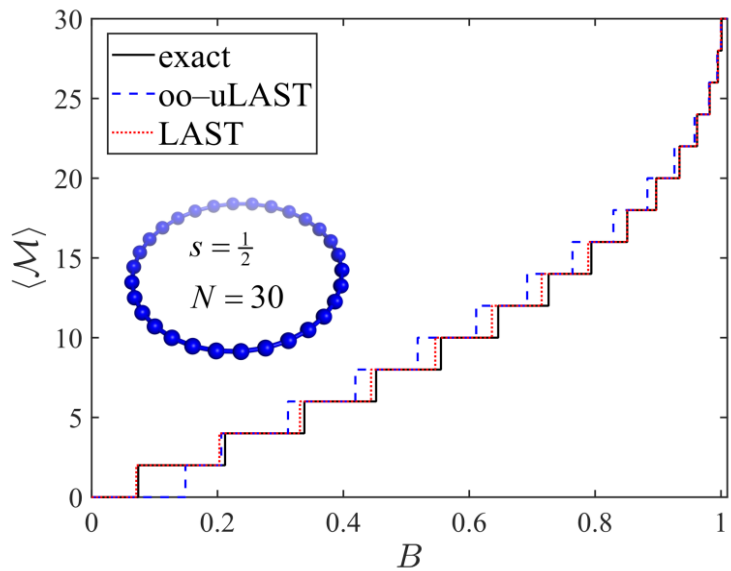

Figure 9: Field-dependent zero-temperature magnetization for an $s=\frac{1}{2}$ ring with $N=30$ sites in the antiferromagnetic Heisenberg model (spheres and connecting lines represent spins and pairwise couplings, respectively).

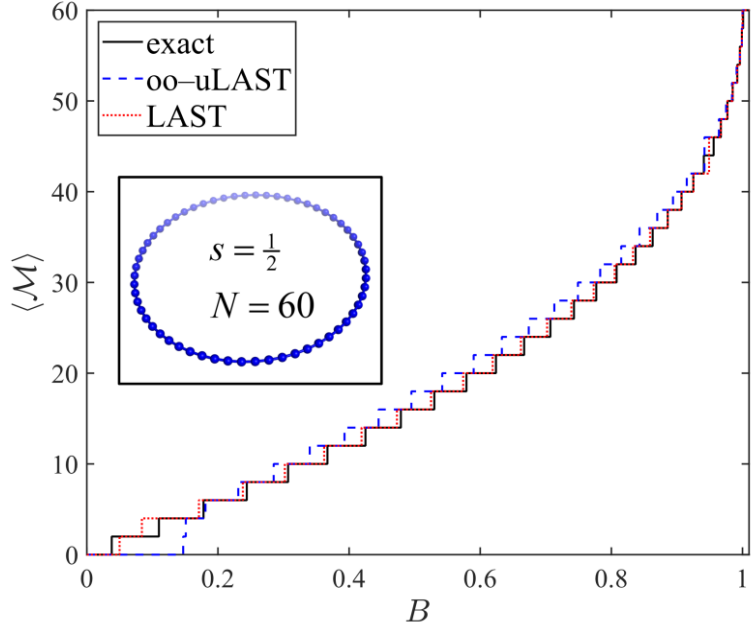

Figure 10: Field-dependent zero-temperature magnetization for an antiferromagnetic $s = \frac{1}{2}$ ring with $N = 60$ sites.

Finally, Figure 11 shows the magnetization staircase for an antiferromagnetic icosidodecahedron. The exact curve was calculated from the ground-state energies in all $M$ sectors reported by Rousochatzakis et al., cf. Table Ib and Figure 5 in Ref. 42. Errors of oo-uLAST as well as LAST are significantly larger than in rings. The most prominent feature is the plateau at $\langle \mathcal{M} \rangle = 10$ (one third of the saturation value). Its width is significantly overestimated by oo-uLAST but captured more accurately by LAST. Away from the plateau region, LAST generally continues to track the exact staircase more closely.

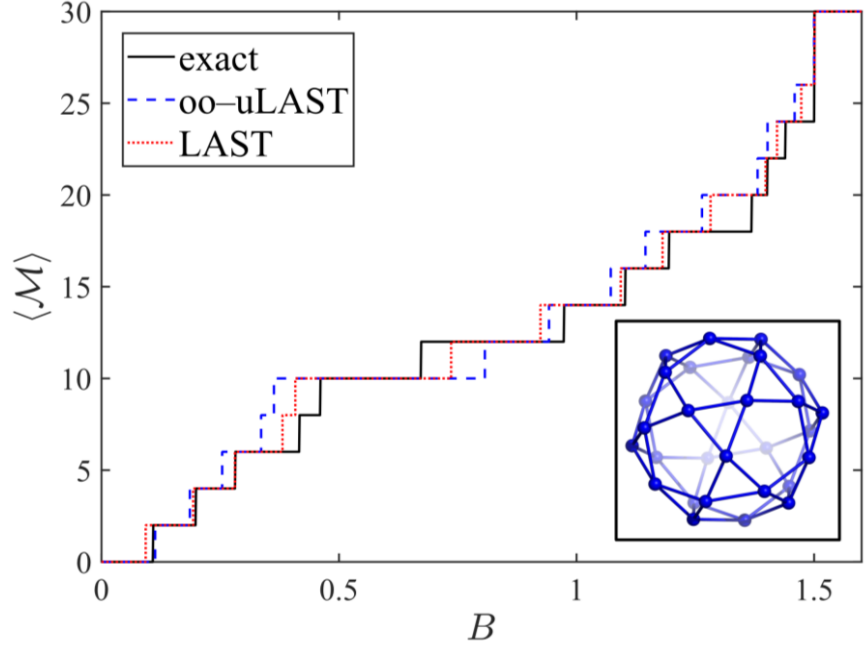

Figure 11: Zero-temperature magnetization as a function of magnetic field for the $s=\frac{1}{2}$ icosidodecahedron in the antiferromagnetic Heisenberg model.

## 4. Summary and Conclusions

We have developed an efficient framework for orbital-optimized unitary Lie-Algebraic Similarity Transformation (oo-uLAST) and its size-extensive non-unitary extension (LAST) for spin Hamiltonians, based on mapping spins to fermions via an extended Jordan–Wigner (EJW) transformation. The analytical energy gradients and Jacobians derived here make the combined optimization of the mean-field state and EJW angles, and the subsequent determination of LAST amplitudes computationally feasible, even for systems with >100 spin-1/2 particles. We suggested an auxiliary-qubit decomposition that makes $s > \frac{1}{2}$ systems amenable to fermionization by EJW. This constitutes a particularly simple scheme, requiring only very minor changes relative to an $s = \frac{1}{2}$ implementation. Although the mapping enlarges the Hilbert space, the procedure is variationally safe: as we prove here, the ground state of a bilinear spin Hamiltonian carries maximal local spins.

Our results confirm that the spin-fermion mapping weakens strong correlation compared to the spin domain. In particular, the variational and size-extensive oo-uLAST approach removes the ambiguity of site ordering and recovers a significant fraction ($\approx$ 97–99 %) of the exact correlation energy for Heisenberg rings up to (and beyond) $N = 30$ sites and for various local spin values (at present up to $s = \frac{5}{2}$).

In summary, the present findings further underline that the strategy of fermionizing finite spin systems and applying simple, cost-effective many-body techniques from electronic-structure theory remains an underexplored, highly promising direction for treating strong correlation that should be pursued further. A natural extension of the present framework would be to replace the single-determinant reference with a compact linear combination of multiple determinants. For instance, an ansatz akin to resonating Hartree–Fock, involving a linear combination of a few simultaneously optimized determinants, could be straightforwardly embedded within the oo-uLAST formalism and would provide a viable starting point for post-mean-field correlation through similarity-transformed LAST. A direct way to construct linear combinations would employ symmetries. However, symmetry operations that are simple in the spin representation often map to complicated, non-local fermionic operators under the JW transformation, making their implementation less straightforward. Some useful exceptions do exist, however, like the spin-flip symmetry in the $M = 0$ sector, which allows to separate even and odd total-spin sectors, or the cyclic point-group symmetry of spin-1/2 rings. In contrast, full adaptation to total-spin eigenstates or the general exploitation of point-group symmetries in more complicated lattices may not always be practical within the fermionic framework.

**Funding.** S.G.T. was supported by the Deutsche Forschungsgemeinschaft (DFG) under Project 535298924. The work at Rice University was supported by the U.S. Department of Energy, Office of Basic Energy Sciences, Computational and Theoretical Chemistry Program under Award DE-SC0001474. G.E.S. acknowledges support as a Welch Foundation Chair (Grant No. C-0036).

**Acknowledgments.** We thank TU Berlin for computational resources.

**Conflicts of Interest.** The authors declare no conflicts of interest.

# Appendix

## A1. Optimization of the mean-field state

Each column of the matrix $\mathbf{C}$ defines a molecular orbital (MO), Eq. (A1):

$$a_i^\dagger = \sum_m C_{mi} c_m^\dagger \ . \tag{A1}$$

There are $(N_{\text{orb}} - N_f) \times N_f$ complex parameters $Z_{vo}$, where $N_{\text{orb}}$ is the number of single-particle basis functions (the number of spin-1/2 particles), and $N_f$ is the number of fermions. The transformation of an initial guess $\mathbf{C}^0$ into $\mathbf{C}$ proceeds through the intermediate $\breve{\mathbf{C}}$ [compare Eqs. (A2) and (A3) to Eqs. 3.45 and 3.47 in Ref. 43],

$$\breve{\mathbf{C}}_o = \mathbf{C}_o^0 + \sum_v Z_{vo} \mathbf{C}_v^0 \ , \tag{A2}$$

$$\breve{\mathbf{C}}_v = \mathbf{C}_v^0 - \sum_o Z_{vo}^* \mathbf{C}_o \ , \tag{A3}$$

where $\breve{\mathbf{C}}_o$ is the $o$-th column of $\breve{\mathbf{C}}$, etc. Eqs. (A4) and (A5) form the orthonormal $\mathbf{C}$,

$$(\mathbf{C}_{\text{occ}})_{lm} = \sum_k (\mathbf{L}^{-1})_{lk} (\breve{\mathbf{C}}_{\text{occ}})_{km} \ , \tag{A4}$$

$$(\mathbf{C}_{\text{virt}})_{lm} = \sum_k (\mathbf{K}^{-1})_{lk} (\breve{\mathbf{C}}_{\text{virt}})_{km} \ , \tag{A5}$$

using lower-triangular matrices $\mathbf{L}$ and $\mathbf{K}$ from the Cholesky decompositions of Eqs. (A6) and (A7):

$$\mathbf{1} + \mathbf{Z}^T \mathbf{Z}^* = \mathbf{L}\mathbf{L}^\dagger \ , \tag{A6}$$

$$\mathbf{1} + \mathbf{Z}^* \mathbf{Z}^T = \mathbf{K}\mathbf{K}^\dagger \ . \tag{A7}$$

Energy and gradient evaluations require single- and two-particle matrix elements between the non-orthogonal determinants $|\Phi\rangle$ and $|\tilde{\Phi}\rangle$. The respective expressions utilize the single-particle transition-density matrix $\boldsymbol{\rho}^{\Phi\tilde{\Phi}}$ with elements $\rho_{kl}^{\Phi\tilde{\Phi}}$,

$$\rho_{kl}^{\Phi\tilde{\Phi}} \equiv \frac{\langle \Phi | c_l^\dagger c_k | \tilde{\Phi} \rangle}{\langle \Phi | \tilde{\Phi} \rangle} = (\tilde{\mathbf{C}}_{\text{occ}} \mathbf{M}^{-1} \mathbf{C}^\dagger)_{kl} \ , \tag{A8}$$

where $\mathbf{M} \equiv \mathbf{C}_{\text{occ}}^\dagger \tilde{\mathbf{C}}_{\text{occ}}$; the overlap is $\langle \Phi | \tilde{\Phi} \rangle = \det(\mathbf{M})$. For a derivation of Eq. (A8), see, e.g., Ref. 43. The two-particle matrix elements are an antisymmetrized product, Eq. (A9); to avoid clutter, we use $\rho_{mk}$ instead of $\rho_{mk}^{\Phi\tilde{\Phi}}$:

$$\frac{\langle\Phi|c_k^\dagger c_l^\dagger c_m c_n|\tilde{\Phi}\rangle}{\langle\Phi|\tilde{\Phi}\rangle} = \begin{vmatrix} \rho_{mk} & \rho_{ml} \\ \rho_{nk} & \rho_{nl} \end{vmatrix} = \rho_{mk}\rho_{nl} - \rho_{ml}\rho_{nk} \ . \tag{A9}$$

For $|\tilde{\Phi}\rangle = |\Phi\rangle$, the single-particle density matrix is $\boldsymbol{\rho} = \mathbf{C}_{\mathrm{occ}}\mathbf{C}_{\mathrm{occ}}^\dagger$ and the evaluation of the $z$-coupling energy is standard, see Eq. (A10),

$$E_z \equiv \langle\Phi|H_z|\Phi\rangle = \mathrm{tr}\left[(\mathbf{t} + \tfrac{1}{2}\boldsymbol{\Gamma})\boldsymbol{\rho}\right] + E_{\mathrm{const}} \ , \tag{A10}$$

with $\boldsymbol{\Gamma}$ defined in Eq. (A11),

$$\Gamma_{kl} = \sum_{mn} [kl\,|\,mn]\rho_{mn} \ . \tag{A11}$$

The $x$- and $y$-coupling contributions are slightly more complicated due to the presence of strings, which act as Thouless rotations.[16] As an example, for $|\tilde{\Phi}\rangle = e^{i\sum_q \theta_{pq} n_q}|\Phi\rangle$, it is straightforward to verify Eq. (A12),

$$\tilde{\mathbf{C}} = \begin{pmatrix} e^{i\theta_{p1}} & 0 & 0 \\ 0 & \ddots & 0 \\ 0 & 0 & e^{i\theta_{pN}} \end{pmatrix} \mathbf{C} \ . \tag{A12}$$

This is indeed a special case of Eq. 6 in Ref. 25. For each interacting pair $\langle m,n\rangle$, we thus consider $|\tilde{\Phi}_{mn}\rangle = \phi_{m(n)}^\dagger \phi_{n(m)}|\Phi\rangle$ and evaluate the respective hopping term, $\tfrac{1}{2} c_m^\dagger c_n$. This overall yields the energy contribution of Eq. (A13),

$$E_{mn}^{xy} \equiv \langle\Phi|s_m^x s_n^x + s_m^y s_n^y|\Phi\rangle = \frac{1}{2}\mathrm{tr}(\mathbf{h}_{mn}\boldsymbol{\rho}^{\Phi\tilde{\Phi}_{mn}})\langle\Phi|\tilde{\Phi}_{mn}\rangle + \mathrm{c.c.} \ , \tag{A13}$$

where $\mathbf{h}^{mn}$, which represents $c_m^\dagger c_n$ in the single-particle basis, has exactly one nonzero entry of $J_{mn}$, located in row $m$ and column $n$.

In the context of PHF for electronic-structure theory, a detailed derivation of the gradient with respect to the Thouless parameters can be found in Ref. 43, which also cites earlier work on gradient-based mean-field calculations. The global gradient $\mathcal{G}$ is defined in terms of energy variations with respect to Thouless rotations applied to the reference $|\Phi^0\rangle$. Its elements $\mathcal{G}_{vo}$ are defined through Eq. (A14), where the real and imaginary parts of $Z_{vo}$ are independent variables.

$$\delta E = \sum_{vo} \frac{\partial E}{\partial Z_{vo}^*} \delta Z_{vo}^* + \text{c.c.} =$$
$$-\sum_{vo} \left( \mathcal{G}_{vo} \delta Z_{vo}^* + \text{c.c.} \right) = \quad (A14)$$
$$-2\sum_{vo} \left[ \text{Re}(\mathcal{G}_{vo}) \text{Re}(\delta Z_{vo}) + \text{Im}(\mathcal{G}_{vo}) \text{Im}(\delta Z_{vo}) \right]$$

We explicitly calculate the local gradient **G** and ultimately transform it into the MO-basis of the initial guess $|\Phi^0\rangle$ to obtain the global gradient $\mathcal{G} = (\mathbf{C}_{\text{virt}}^0)^\dagger \mathbf{G} \mathbf{C}_{\text{occ}}^0$, see Eq. (A15),

$$\mathcal{G} = (\mathbf{C}_{\text{virt}}^0)^\dagger \mathbf{G} \mathbf{C}_{\text{occ}}^0 . \quad (A15)$$

The respective contributions of the decomposition in Eq. (A16),

$$\mathbf{G} = \mathbf{G}_z + \sum_{m<n} \mathbf{G}_{mn}^{xy} , \quad (A16)$$

are given in Eqs. (A17) and (A18).

$$(\mathbf{G}_z)_{vo} = -\left[ (\boldsymbol{\rho}-\mathbf{1})(\mathbf{t}+\boldsymbol{\Gamma}) - (E_z - E_{\text{const}}) \right]\boldsymbol{\rho} \quad (A17)$$

$$\mathbf{G}_{mn}^{xy} = -\frac{1}{2}\left[ (\boldsymbol{\rho}^{\Phi\tilde{\Phi}_{mn}} - \mathbf{1})\mathbf{h}_{mn} - \text{tr}(\mathbf{h}_{mn}\boldsymbol{\rho}^{\Phi\tilde{\Phi}_{mn}}) \right]\boldsymbol{\rho}^{\Phi\tilde{\Phi}_{mn}} \langle \Phi|\tilde{\Phi}_{mn}\rangle + \text{c.c.} \quad (A18)$$

### A2. uLAST gradient

Here, to optimize the uLAST parameters $\theta_{pq}$ efficiently, we derive the gradient $\vartheta$ defined in Eq. (A19):

$$\vartheta_{pq} \equiv \frac{\partial \langle \Phi | H(\boldsymbol{\theta}) | \Phi \rangle}{\partial \theta_{pq}} = \langle \Phi | \frac{\partial H(\boldsymbol{\theta})}{\partial \theta_{pq}} | \Phi \rangle . \quad (A19)$$

The derivative of the combined $x$- and $y$-coupling terms for a pair $\langle m,n \rangle$ is considered in Eq. (A20).

$$\frac{\partial}{\partial \theta_{pq}} \left( s_m^x s_n^x + s_m^y s_n^y \right) = \frac{\partial}{\partial \theta_{pq}} [\tfrac{1}{2} c_m^\dagger c_n \phi_{m(n)}^\dagger \phi_{n(m)} + \text{h.c.}] =$$
$$\frac{1}{2} c_m^\dagger c_n \left[ \phi_{n(m)} \frac{\partial \phi_{m(n)}^\dagger}{\partial \theta_{pq}} + \phi_{m(n)}^\dagger \frac{\partial \phi_{n(m)}}{\partial \theta_{pq}} \right] + \text{h.c.} \quad (A20)$$

For one term on the right-hand side of Eq. (A20), the calculation is exemplarily detailed in Eq. (A21), where $\theta_{qp} = \theta_{pq} + \pi$ for $p < q$.

$$\frac{\partial \phi_{n(m)}}{\partial \theta_{pq}} = \frac{\partial e^{-i\sum_{r\neq m}\theta_{nr}n_r}}{\partial \theta_{pq}} = -i\frac{\partial\left[\theta_{nq}n_q(1-\delta_{mq})+\theta_{np}n_p(1-\delta_{mp})\right]}{\partial \theta_{pq}}\phi_{n(m)} =$$
$$-i\frac{\partial\left[\theta_{nq}n_q(1-\delta_{mq})+(\theta_{pn}\pm\pi)n_p(1-\delta_{mp})\right]}{\partial \theta_{pq}}\phi_{n(m)} = \tag{A21}$$
$$-i\left[\delta_{np}(1-\delta_{mq})n_q+\delta_{nq}(1-\delta_{mp})n_p\right]\phi_{n(m)}$$

A similar calculation yields Eq. (A22),

$$\frac{\partial \phi_{m(n)}^\dagger}{\partial \theta_{pq}} = i\left[\delta_{mp}(1-\delta_{nq})n_q+\delta_{mq}(1-\delta_{np})n_p\right]\phi_{m(n)} , \tag{A22}$$

and we overall obtain Eq. (A23).

$$\frac{\partial}{\partial \theta_{pq}}\left[\tfrac{1}{2}c_m^\dagger c_n \phi_{m(n)}^\dagger \phi_{n(m)} + \text{h.c.}\right] = \frac{i}{2}\Big\{c_m^\dagger c_n\left[\delta_{mp}(1-\delta_{nq})-\delta_{np}(1-\delta_{mq})\right]n_q$$
$$+\left[\delta_{mq}(1-\delta_{np})-\delta_{nq}(1-\delta_{mp})\right]n_p\Big\}\phi_{m(n)}^\dagger \phi_{n(m)} + \text{h.c.} \tag{A23}$$

Establishing normal order through Eq. (A24),

$$c_m^\dagger c_n n_r = c_m^\dagger c_n c_r^\dagger c_r = c_m^\dagger(\delta_{nr}-c_r^\dagger c_n)c_r = \delta_{nr}c_m^\dagger c_r - c_m^\dagger c_r^\dagger c_n c_r , \tag{A24}$$

gives Eq. (A25).

$$\frac{\partial}{\partial \theta_{pq}}\left[\tfrac{1}{2}c_m^\dagger c_n \phi_{m(n)}^\dagger \phi_{n(m)} + \text{h.c.}\right] =$$
$$\frac{i}{2}\Big\{\left[\delta_{mp}(1-\delta_{nq})-\delta_{np}(1-\delta_{mq})\right](\delta_{nq}c_m^\dagger c_q - c_m^\dagger c_q^\dagger c_n c_q) \tag{A25}$$
$$+\left[\delta_{mq}(1-\delta_{np})-\delta_{nq}(1-\delta_{mp})\right](\delta_{np}c_m^\dagger c_p - c_m^\dagger c_p^\dagger c_n c_p)\Big\}\phi_{m(n)}^\dagger \phi_{n(m)} + \text{h.c.}$$

The single-particle contribution in Eq. (A25) obviously vanishes, which yields Eq. (A26).

$$\frac{\partial(\mathbf{s}_m \cdot \mathbf{s}_n)}{\partial \theta_{pq}} = -\frac{i}{2}\Big\{\left[\delta_{mp}(1-\delta_{nq})-\delta_{np}(1-\delta_{mq})\right]c_m^\dagger c_q^\dagger c_n c_q$$
$$+\left[\delta_{mq}(1-\delta_{np})-\delta_{nq}(1-\delta_{mp})\right]c_m^\dagger c_p^\dagger c_n c_p\Big\}\phi_{m(n)}^\dagger \phi_{n(m)} + \text{h.c.} \tag{A26}$$

$E$, $\mathbf{C}^0$, $\mathcal{G}$ and $\vartheta$ are passed to `fminunc`, a `Matlab` function for unconstrained minimization. We separate terms as in Eq. (A14), because `fminunc` requires real optimization parameters.

### A3. LAST residuals and Jacobian

A string operator resulting from a general LAST (including a uLAST component) scales the orbital coefficients non-uniformly, see Eq. (A27) for $|\tilde{\Phi}\rangle = e^{\sum_q (\alpha_{pq}+i\theta_{pq})n_q}|\Phi\rangle$.

$$\tilde{\mathbf{C}}_{occ} = \begin{pmatrix} e^{\alpha_{p1}+i\theta_{p1}} & 0 & 0 \\ 0 & \ddots & 0 \\ 0 & 0 & e^{\alpha_{pN}+i\theta_{pN}} \end{pmatrix} \mathbf{C}_{occ} \quad (A27)$$

While the columns of $\tilde{\mathbf{C}}_{occ}$ are neither normalized nor orthogonal, $|\tilde{\Phi}\rangle$ still represents a single Slater determinant up to a prefactor.[25] Importantly, Eqs. (A8) and (A9) for the single- and two-particle matrix elements remain valid. Eq. (A28) shows the form of *x*- and *y*-coupling contributions to the residuals,

$$\langle \Phi | n_p n_q c_m^\dagger c_n \phi_{m(n)}^\dagger \phi_{n(m)} | \Phi \rangle , \quad (A28)$$

and Eq. (A29) establishes normal order,

$$n_p n_q c_m^\dagger c_n = c_p^\dagger c_q^\dagger c_m^\dagger c_q c_p c_n - \delta_{mq} c_p^\dagger c_q^\dagger c_p c_n + \delta_{mp} c_p^\dagger c_q^\dagger c_q c_n . \quad (A29)$$

The three-particle term is evaluated based on the general Eq. (A30):

$$\frac{\langle \Phi | c_k^\dagger c_l^\dagger c_m^\dagger c_n c_o c_p | \tilde{\Phi} \rangle}{\langle \Phi | \tilde{\Phi} \rangle} = - \begin{vmatrix} \rho_{nk} & \rho_{nl} & \rho_{nm} \\ \rho_{ok} & \rho_{ol} & \rho_{om} \\ \rho_{pk} & \rho_{pl} & \rho_{pm} \end{vmatrix} . \quad (A30)$$

The single-particle terms from *z*-coupling contribute terms of the form $\langle \Phi | n_p n_q n_m | \Phi \rangle$ to the residuals, which can be evaluated similarly. For the two-particle contribution from *z*-coupling, however, we must append another hopping operator to the right of Eq. (A31).

$$\begin{aligned} n_p n_q c_m^\dagger c_n c_k^\dagger c_l &= c_p^\dagger c_q^\dagger c_m^\dagger c_q c_p c_n c_k^\dagger c_l - \delta_{mq} c_p^\dagger c_q^\dagger c_p c_n c_k^\dagger c_l + \delta_{mp} c_p^\dagger c_q^\dagger c_q c_n c_k^\dagger c_l = \\ & c_p^\dagger c_q^\dagger c_m^\dagger \delta_{kn} c_q c_p c_l - c_p^\dagger c_q^\dagger c_m^\dagger \delta_{kp} c_q c_n c_l + c_p^\dagger c_q^\dagger c_m^\dagger \delta_{kq} c_p c_n c_l - c_p^\dagger c_q^\dagger c_m^\dagger c_k^\dagger c_q c_p c_n c_l \\ & -\delta_{mq} (\delta_{kn} c_p^\dagger c_q^\dagger c_p c_l - \delta_{kp} c_p^\dagger c_q^\dagger c_n c_l + c_p^\dagger c_q^\dagger c_k^\dagger c_p c_n c_l ) \\ & +\delta_{mp} (\delta_{kn} c_p^\dagger c_q^\dagger c_q c_l - \delta_{kq} c_p^\dagger c_q^\dagger c_n c_l + c_p^\dagger c_q^\dagger c_k^\dagger c_q c_n c_l ) \end{aligned} \quad (A31)$$

Eq. (A31) contains a four-particle term, which is evaluated based on Eq. (A32).

$$\frac{\langle \Phi | c_k^\dagger c_l^\dagger c_m^\dagger c_n^\dagger c_o c_p c_q c_r | \tilde{\Phi} \rangle}{\langle \Phi | \tilde{\Phi} \rangle} = \begin{vmatrix} \rho_{ok} & \rho_{ol} & \rho_{om} & \rho_{on} \\ \rho_{pk} & \rho_{pl} & \rho_{pm} & \rho_{pn} \\ \rho_{qk} & \rho_{ql} & \rho_{qm} & \rho_{qn} \\ \rho_{rk} & \rho_{rl} & \rho_{rm} & \rho_{rn} \end{vmatrix} \quad (A32)$$

The calculation of the Jacobian, Eq. (26) in the main text, involves the derivative of $\bar{H}$ with respect to $\alpha_{pq}$, Eq. (A33). The steps to arrive at this result are not given here, because they are very similar to the derivation of the uLAST gradient (see previous section).

$$\frac{\partial \bar{H}}{\partial \alpha_{pq}} = -\frac{1}{2} \sum_{m<n} J_{mn} \left\{ \left[ \delta_{mp}(1-\delta_{nq}) + \delta_{np}(1-\delta_{mq}) \right] c_m^\dagger c_q^\dagger c_n c_q \right.$$
$$\left. + \left[ \delta_{mq}(1-\delta_{np}) + \delta_{nq}(1-\delta_{mp}) \right] c_m^\dagger c_p^\dagger c_n c_p \right\} \phi_{m(n)}^\dagger \phi_{n(m)} + \text{h.c.} \quad (A33)$$

## A4. Proof: Auxiliary qubits form maximal local spins in the ground state of a bilinear Hamiltonian

Consider a general bilinear spin Hamiltonian, Eq. (A34), where the local spin-quantum numbers $s_p$ may be different for different sites. The pairwise coupling is parametrized by $3\times 3$ matrices $\mathbf{D}_{mn}$ (Cartesian rank-2 tensors) with arbitrary values of the real components $D_{mn}^{\alpha\beta}$,

$$H = \sum_{m<n} \mathbf{s}_m \cdot \mathbf{D}_{mn} \cdot \mathbf{s}_n = \sum_{m<n} \sum_{\alpha,\beta=x,y,z} D_{mn}^{\alpha\beta} s_m^\alpha s_n^\beta \ . \quad (A34)$$

In general, $H$ does not conserve $S_z$, which is a case not directly relevant to the mean-field calculations performed in this work. However, the *XXZ* model, which conserves $S_z$, is a special case of Eq. (A34) where $\mathbf{D}_{mn} = \text{diag}(J_{mn}, J_{mn}, \Delta_{mn})$ for all pairs $\langle m,n \rangle$.

We extend the Hilbert space as shown in Eq. (A35),[d]

$$\mathbf{s}_p \to \tilde{\mathbf{s}}_p = \sum_{a=1}^{2s_p} \boldsymbol{\kappa}_{p,a} \ , \quad (A35)$$

by introducing $2s_p$ spin-1/2 auxiliaries at each site. In the following, we prove that the ground state of $\tilde{H}$, Eq. (A36),

$$H \to \tilde{H} = \sum_{m<n} \tilde{\mathbf{s}}_m \cdot \mathbf{D}_{mn} \cdot \tilde{\mathbf{s}}_n \ , \quad (A36)$$

belongs to the space of the physical Hamiltonian $H$, i.e., all on-site spins are maximal. We are not aware of any previous proof of this proposition in the literature.

Note first that, because local spin is a good quantum number, i.e., $[\tilde{H}, \tilde{\mathbf{s}}_p^2] = 0$ for each $p$, the Hilbert space splits into sectors labeled by the set of on-site spins $(t_1, t_2, ..., t_N)$, such that $\tilde{\mathbf{s}}_p^2 = t_p(t_p+1)$. $\tilde{H}$ can be diagonalized independently in each sector. In the following, we focus on a site $p$. As shown in Eq. (A37), we separate $\tilde{H}$ into a term $\tilde{H}_{\text{rest}}$ that is independent of $\tilde{\mathbf{s}}_p$,

---

[d] As apparent, certain operators carry a tilde in the auxiliary-qubit space. To avoid clutter, we do not use tildes on operators that are not defined (or trivial) in the original space of $H$.

Eq. (A38), and a term $\tilde{\mathbf{s}}_p \cdot \tilde{\mathbf{B}}$ for the interaction of $\tilde{\mathbf{s}}_p$ with its environment; the superscript $T$ in Eq. (A39) denotes matrix transposition.

$$\tilde{H} = \tilde{H}_{\text{rest}} + \sum_{m>p} \tilde{\mathbf{s}}_p \cdot \mathbf{D}_{pm} \cdot \tilde{\mathbf{s}}_m + \sum_{n<p} \tilde{\mathbf{s}}_n \cdot \mathbf{D}_{np} \cdot \tilde{\mathbf{s}}_p = \tilde{H}_{\text{rest}} + \tilde{\mathbf{s}}_p \cdot \tilde{\mathbf{B}} \tag{A37}$$

$$\tilde{H}_{\text{rest}} = \sum_{\substack{m<n \\ m,n \neq p}} \tilde{\mathbf{s}}_m \cdot \mathbf{D}_{mn} \cdot \tilde{\mathbf{s}}_n \tag{A38}$$

$$\tilde{\mathbf{B}} = \sum_{m>p} \mathbf{D}_{pm} \cdot \tilde{\mathbf{s}}_m + \sum_{n<p} \mathbf{D}_{np}^T \cdot \tilde{\mathbf{s}}_n \tag{A39}$$

The ground state in a sector $(t_1, t_2, \ldots, t_N)$ is degenerate with respect to the linearly independent ways of coupling the auxiliaries into their respective local spins $(t_1, t_2, \ldots, t_N)$.[e] Therefore, we may choose – without loss of generality – a ground state $|\Psi_{t_p}\rangle$ that has $(s_p - t_p)$ localized singlets at site $p$. We assume that $\kappa_{p,1}$ and $\kappa_{p,2}$ form a singlet if $s_p - t_p \geq 1$, $\kappa_{p,3}$ and $\kappa_{p,4}$ form a second singlet if $s_p - t_p \geq 2$, etc. The remaining $2t_p$ qubits are coupled into a total spin $t_p$. Thus, $|\Psi_{t_p}\rangle$ comprises a product of singlets at $p$, see Eq. (A40),

$$|\Psi_{t_p}\rangle = |0\rangle_{(1,2)} |0\rangle_{(3,4)} |0\rangle_{(5,6)} \ldots |\psi_{t_p}\rangle \tag{A40}$$

where $|\psi_{t_p}\rangle$ represents the remaining $2t_p$ auxiliaries as well as the other sites (in general, these $2t_p$ qubits are entangled with the other sites). When excluding a pair (1,2) from the product, we use the notation of Eq. (A41):

$$|_{(1,2)}\Psi_{t_p}\rangle \equiv |0\rangle_{(3,4)} |0\rangle_{(5,6)} \ldots |\psi_{t_p}\rangle . \tag{A41}$$

We shall prove the maximal local-spin conjecture inductively. In the basis step, we consider a sector with minimal spin at $p$, i.e., $(t_p)_{\min} = 0$ or $(t_p)_{\min} = \frac{1}{2}$, depending on whether $2s_p$ is even or odd, respectively. The other values $\{t_i\}$, $i \neq p$, remain fixed throughout. In the following, we denote $(t_p)_{\min}$ by $t_{\min}$ and $t_p$ by $t$. Our aim is to show that the ground-state energy $E_0^{t_{\min}} \equiv \langle \Psi_{t_{\min}} | \tilde{H} | \Psi_{t_{\min}} \rangle$ is an upper bound to the ground state $E_0^{t_{\min}+1}$ of the respective sector where the spin at $p$ is increased by one unit,

---

[e] Additional degeneracies may result from non-Abelian symmetries of the physical Hamiltonian $H$ (e.g., spin-rotational symmetry in the Heisenberg model) but are of no concern here.

$$E_0^{t_{\min}+1} \leq E_0^{t_{\min}} . \tag{A42}$$

By the variational principle, Eq. (A42) holds if there exists a trial state $\left|\Phi_{t_{\min}+1}\right\rangle$ with $(t_{\min}+1)$ that satisfies $E_{\text{trial}} \leq E_0^{t_{\min}}$. In the following, we transform $\left|\Psi_{t_{\min}}\right\rangle$ into $\left|\Phi_{t_{\min}+1}\right\rangle$ and show that the supposed inequality holds. To this end, the singlet $|0\rangle$ formed by $\boldsymbol{\kappa}_{p,1}$ and $\boldsymbol{\kappa}_{p,2}$ is first decoupled into a triplet $|1\rangle$ through a unitary operator $R$, i.e., $R|0\rangle = |1\rangle$. A general triplet state is a superposition of three possible projections on a spatial direction defined by a unit vector $\mathbf{u}$, i.e., $|1,-1\rangle_{\mathbf{u}}$, $|1,0\rangle_{\mathbf{u}}$ and $|1,+1\rangle_{\mathbf{u}}$. A particular instance of $R$ is $R_{\mathbf{u}}$, which releases $|0\rangle$ into $|1,-1\rangle_{\mathbf{u}}$, see Eqs. (A43) and (A44),

$$R_{\mathbf{u}}|0\rangle = |1,-1\rangle_{\mathbf{u}} , \tag{A43}$$

$$_{\mathbf{u}}\langle 1,-1|(\boldsymbol{\kappa}_{p,1}+\boldsymbol{\kappa}_{p,2})|1,-1\rangle_{\mathbf{u}} = -\mathbf{u} . \tag{A44}$$

A normalized trial state $|\theta_t\rangle$ is constructed from $|\Psi_t\rangle$ by releasing a singlet pair $(a,b)$,

$$|\theta_t\rangle \equiv R_{(a,b),\mathbf{u}}|\Psi_t\rangle = |1,-1\rangle_{(a,b),\mathbf{u}} |_{(a,b)}\Psi_t\rangle . \tag{A45}$$

Using Eq. (A44) and the definition of Eq. (A45), $E_{\text{trial}}$ is evaluated in Eq. (A46),

$$\begin{aligned} E_{\text{trial}} &= \left\langle \theta_{t_{\min}} \left| \tilde{H} \right| \theta_{t_{\min}} \right\rangle \\ &= \left\langle \Psi_{t_{\min}} \left| \tilde{H} \right| \Psi_{t_{\min}} \right\rangle + {}_{\mathbf{u}}\langle 1,-1|(\boldsymbol{\kappa}_{p,1}+\boldsymbol{\kappa}_{p,2})|1,-1\rangle_{\mathbf{u}} \cdot \left\langle \Psi_{t_{\min}} \left| \tilde{\mathbf{B}} \right| \Psi_{t_{\min}} \right\rangle , \\ &= E_0^{t_{\min}} - \mathbf{u} \cdot \langle \tilde{\mathbf{B}} \rangle \end{aligned} \tag{A46}$$

where $\langle \tilde{\mathbf{B}} \rangle \equiv \langle \Psi_{t_{\min}} | \tilde{\mathbf{B}} | \Psi_{t_{\min}} \rangle$. A direction $\mathbf{u}$ can always be chosen such that $E_{\text{trial}} \leq E_0^{t_{\min}}$, and, unless $\langle \tilde{\mathbf{B}} \rangle = 0$, a strict inequality can be obtained, $E_{\text{trial}} < E_0^{t_{\min}}$. If $t_{\min} = 0$, then $|\theta_{t_{\min}}\rangle$ has a definite $t_p = 1$, and the induction basis has been established.

On the other hand, if $t_{\min} = \frac{1}{2}$, then $|\theta_{t_{\min}}\rangle$ is a superposition of $t = \frac{1}{2}$ and $t = \frac{3}{2}$ states. To analyze this case, consider that any state $|\Xi\rangle$ can be expressed as a linear combination with respect to the on-site spin at $p$, see Eq. (A47):

$$|\Xi\rangle = \sum_t c_t |\Xi_t\rangle . \tag{A47}$$

The individual contributions are picked out by a (Hermitian and idempotent) spin-projection operator $P^{(t)}$, Eq. (A48),

$$P^{(t)}|\Xi\rangle = c_t|\Xi_t\rangle \ . \tag{A48}$$

Note that the projector can be expressed in the Löwdin[44] form of Eq. (A49),

$$P^{(t)} = \prod_{l \neq t} \frac{\tilde{\mathbf{s}}_p^2 - l(l+1)}{t(t+1) - l(l+1)} \ , \tag{A49}$$

but such a specific expression is not needed here. For a Hamiltonian with local spin as a good quantum number, $[\tilde{H}, \tilde{\mathbf{s}}_p^2] = 0$ (thereby $[\tilde{H}, P^{(t)}] = 0$), the energy is the probability-weighted sum of Eq. (A50),

$$E = \langle \Xi|\tilde{H}|\Xi\rangle = \sum_t \frac{\langle \Xi|\tilde{H}P^{(t)}|\Xi\rangle}{\langle \Xi|P^{(t)}|\Xi\rangle} = \sum_t |c_t|^2 \langle \Xi_t|\tilde{H}|\Xi_t\rangle = \sum_t |c_t|^2 E_t \ . \tag{A50}$$

In the specific case of $|\theta_{t_{\min}}\rangle$ with $t_{\min} = \frac{1}{2}$, $E_{\text{trial}}$ of Eq. (A46) is the weighted average $E_{\text{trial}} = |c_{1/2}|^2 E_{1/2} + |c_{3/2}|^2 E_{3/2}$, with $|c_{1/2}|^2 + |c_{3/2}|^2 = 1$. But as $E_0^{t_{\min}} \leq E_{1/2}$, we obtain $E_{3/2} \leq E_0^{t_{\min}}$ for the trial state $|\Phi_{t_{\min}+1}\rangle = P^{(3/2)}|\theta_{t_{\min}}\rangle$.

In summary, in both cases ($t_{\min} = 0$ or $t_{\min} = \frac{1}{2}$), $|\Phi_{t_{\min}+1}\rangle \equiv P^{(t_{\min}+1)}|\theta_{t_{\min}}\rangle$ yields $E_{t_{\min}+1} \leq E_0^{t_{\min}}$ (or $E_{t_{\min}+1} < E_0^{t_{\min}}$ if $\langle \tilde{\mathbf{B}}\rangle \neq 0$); thus, by the variational principle, $E_0^{t_{\min}+1} \leq E_{\text{trial}}$. This proves that increasing the spin at $p$ from $t_{\min}$ to $(t_{\min}+1)$ does not increase the ground-state energy.

We proceed inductively and consider a ground state $|\Psi_t\rangle$ with $t_{\min} < t < s_p$. Like in the previous derivation, a singlet $(a,b)$ in $|\Psi_t\rangle$ can be decoupled to yield a favorable interaction between the triplet and the environment [cf. Eq. (A51), where $\langle \tilde{\mathbf{B}}\rangle = \langle \Psi_t|\tilde{\mathbf{B}}|\Psi_t\rangle$]. The choice of an appropriate $\mathbf{u}$ for $|\theta_t\rangle = R_{(a,b),\mathbf{u}}|\Psi_t\rangle$ does not depend on the specific pair $(a,b)$.

$$E_{\text{trial}} = \langle \theta_t|\tilde{H}|\theta_t\rangle = E_0^t - \mathbf{u} \cdot \langle \tilde{\mathbf{B}}\rangle \leq E_0^t \tag{A51}$$

Now, $|\theta_t\rangle$ is a superposition of $(t-1)$, $t$, and $(t+1)$ states, Eq. (A52),

$$E_{\text{trial}} = \langle \theta_t|\tilde{H}|\theta_t\rangle = |c_{t-1}|^2 E_{t-1} + |c_t|^2 E_t + |c_{t+1}|^2 E_{t+1} \ , \tag{A52}$$

with $|c_{t-1}|^2 + |c_t|^2 + |c_{t+1}|^2 = 1$. Inducing from the base case yields the first inequality in (A53),

$$E_{\text{trial}} \geq |c_{t-1}|^2 E_0^t + |c_t|^2 E_t + |c_{t+1}|^2 E_{t+1}$$
$$\geq \left(|c_{t-1}|^2 + |c_t|^2\right) E_0^t + |c_{t+1}|^2 E_{t+1} \quad , \tag{A53}$$

and combining Eqs. (A51) and (A53) yields Eq. (A54):

$$E_0^t \geq \left(|c_{t-1}|^2 + |c_t|^2\right) E_0^t + |c_{t+1}|^2 E_{t+1}$$
$$\Leftrightarrow E_0^t \geq E_{t+1} \quad . \tag{A54}$$

Thus, the (unnormalized) trial state $|\Phi_{t+1}\rangle \equiv P^{(t+1)}|\theta_t\rangle$ does not have a higher energy than $|\Psi_t\rangle$, so that $E_0^{t+1} \leq E_0^t$. This proves the monotonicity with respect to increasing the on-site spin at $p$. Iterating over the sites then affords the result that all on-site spins are maximal in the ground state.

If the Hamiltonian conserves $S_z$, then a ground state $|\Psi_t^M\rangle$ in a magnetization sector $M$ can be similarly used to construct a trial state $|\Phi_{t+1}^M\rangle$, which remains in the same $M$-sector when choosing a projection of 0 of the triplet on the $z$-axis ($\mathbf{u} = \mathbf{z}$). Although such a triplet state does not interact with the environment $\tilde{\mathbf{B}}$, we can still conclude that the ground state can be chosen to have maximal local spins in each $M$-sector.

Although not relevant for the present work, we may speculate that the maximal local-spin principle might more generally apply to the ground state in each symmetry sector $\Gamma$ of a bilinear Hamiltonian, e.g., in each total-spin and magnetization sector $(S,M)$ or for combinations of total-spin and point-group symmetry in the isotropic Heisenberg model,[45,46,47,48] or for symmetry groups comprising combined spin rotations and spin permutations in anisotropic systems with spatial symmetry.[49,50] Besides, we have not been concerned with identifying the most general class of spin Hamiltonians (beyond bilinear) where the maximal local-spin property may hold.

# References

(1) Bencini, A.; Gatteschi, D. *Electron Paramagnetic Resonance of Exchange Coupled Systems*; Springer, Berlin, 1990.

(2) Mattis, D. C. *The Theory of Magnetism I: Statics and Dynamics*; Springer, Berlin, 1988.

(3) Nolting, W.; Ramakanth, A. *Quantum Theory of Magnetism*; Springer Science & Business Media, 2009.

(4) Sandvik, A. W. Computational Studies of Quantum Spin Systems. In *AIP Conference Proceedings*; 2010; Vol. 1297, pp 135–338.

(5) Schollwöck, U.; Richter, J.; Farnell, D. J. J.; Bishop, R. F. *Quantum Magnetism*; Springer, 2008.

(6) Schnack, J.; Ummethum, J. Advanced Quantum Methods for the Largest Magnetic Molecules. *Polyhedron* **2013**, *66*, 28–33.

(7) Jiménez-Hoyos, C. A.; Henderson, T. M.; Tsuchimochi, T.; Scuseria, G. E. Projected Hartree–Fock Theory. *J. Chem. Phys.* **2012**, *136* (16), 164109.

(8) Ghassemi Tabrizi, S.; Jiménez-Hoyos, C. A. Ground States of Heisenberg Spin Clusters from Projected Hartree–Fock Theory. *Phys. Rev. B* **2022**, *105* (3), 35147.

(9) Ghassemi Tabrizi, S.; Jiménez-Hoyos, C. A. Ground States of Heisenberg Spin Clusters from a Cluster-Based Projected Hartree–Fock Approach. *Condens. Matter* **2023**, *8* (1), 18.

(10) Papastathopoulos-Katsaros, A.; Henderson, T. M.; Scuseria, G. E. Symmetry-Projected Cluster Mean-Field Theory Applied to Spin Systems. *J. Chem. Phys.* **2023**, *159* (8).

(11) Jiménez-Hoyos, C. A.; Rodríguez-Guzmán, R.; Scuseria, G. E. Multi-Component Symmetry-Projected Approach for Molecular Ground State Correlations. *J. Chem. Phys.* **2013**, *139* (20).

(12) Li, S. Block-Correlated Coupled Cluster Theory: The General Formulation and Its Application to the Antiferromagnetic Heisenberg Model. *J. Chem. Phys.* **2004**, *120* (11), 5017–5026.

(13) Sanchez-Marin, J.; Malrieu, J. P.; Maynau, D. Approximate Solutions of Heisenberg Hamiltonians. *Int. J. Quantum Chem.* **1987**, *31* (6), 903–925.

(14) Papastathopoulos-Katsaros, A.; Jiménez-Hoyos, C. A.; Henderson, T. M.; Scuseria, G. E. Coupled Cluster and Perturbation Theories Based on a Cluster Mean-Field Reference Applied to Strongly Correlated Spin Systems. *J. Chem. Theory Comput.* **2022**, *18* (7), 4293–4303.

(15) Jordan, P.; Wigner, E. Über das Paulische Äquivalenzverbot. *Zeitschrift für Phys.* **1928**, *47* (9), 631–651.

(16) Henderson, T. M.; Chen, G. P.; Scuseria, G. E. Strong-Weak Duality via Jordan–Wigner Transformation: Using Fermionic Methods for Strongly Correlated SU(2) Spin Systems. *J. Chem. Phys.* **2022**, *157* (19), 194114.

(17) Lieb, E.; Schultz, T.; Mattis, D. Two Soluble Models of an Antiferromagnetic Chain. *Ann. Phys.* **1961**, *16* (3), 407–466.

(18) Thouless, D. J. Stability Conditions and Nuclear Rotations in the Hartree–Fock Theory. *Nucl. Phys.* **1960**, *21*, 225–232.

(19) Dai, X.; Su, Z. Mean-Field Theory for the Spin-Ladder System. *Phys. Rev. B* **1998**, *57* (2), 964.

(20) Nunner, T. S.; Kopp, T. Jordan–Wigner Approach to Dynamic Correlations in Spin Ladders. *Phys. Rev. B* **2004**, *69* (10), 104419.

(21) Verkholyak, T.; Strečka, J.; Jaščur, M.; Richter, J. Magnetic Properties of the Quantum Spin-*XX* Diamond Chain: The Jordan–Wigner Approach. *Eur. Phys. J. B* **2011**, *80* (4), 433–444.

(22) Azzouz, M. Interchain-Coupling Effect on the One-Dimensional Spin-1/2 Antiferromagnetic Heisenberg Model. *Phys. Rev. B* **1993**, *48* (9), 6136.

(23) Wang, Y. R. Ground State of the Two-Dimensional Antiferromagnetic Heisenberg Model Studied Using an Extended Wigner–Jordan Transformation. *Phys. Rev. B* **1991**, *43* (4), 3786.

(24) Henderson, T. M.; Gao, F.; Scuseria, G. E. Restoring Permutational Invariance in the Jordan–Wigner Transformation. *Mol. Phys.* **2023**, e2254857.

(25) Wahlen-Strothman, J. M.; Jiménez-Hoyos, C. A.; Henderson, T. M.; Scuseria, G. E. Lie Algebraic Similarity Transformed Hamiltonians for Lattice Model Systems. *Phys. Rev. B* **2015**, *91* (4), 41114.

(26) Fukutome, H.; Yamamura, M.; Nishiyama, S. A New Fermion Many-Body Theory Based on the SO(2$N$+1) Lie Algebra of the Fermion Operators. *Prog. Theor. Phys.* **1977**, *57* (5), 1554–1571.

(27) Henderson, T. M.; Ghassemi Tabrizi, S.; Chen, G. P.; Scuseria, G. E. Hartree–Fock–Bogoliubov Theory for Number-Parity-Violating Fermionic Hamiltonians. *J. Chem. Phys.* **2024**, *160* (6).

(28) Blaizot, J.-P.; Ripka, G. *Quantum Theory of Finite Systems*; MIT Press, Cambridge, MA, 1986.

(29) Affleck, I.; Kennedy, T.; Lieb, E. H.; Tasaki, H. Valence Bond Ground States in Isotropic Quantum Antiferromagnets. *Commun. Math. Phys.* **1988**, *115* (3), 477–528.

(30) Ghassemi Tabrizi, S.; Kühne, T. D. Projective Spin Adaptation for the Exact Diagonalization of Isotropic Spin Clusters. *Magnetism* **2024**, *4* (4), 332–347.

(31) Solyom, J.; Timonen, J. Spin-1 Heisenberg Chain and the One-Dimensional Fermion Gas. *Phys. Rev. B* **1989**, *40* (10), 7150.

(32) Kim, E. H.; Sólyom, J. Opening of the Haldane Gap in Anisotropic Two-and Four-Leg Spin Ladders. *Phys. Rev. B* **1999**, *60* (22), 15230.

(33) Batista, C. D.; Ortiz, G. Generalized Jordan–Wigner Transformations. *Phys. Rev. Lett.* **2001**, *86* (6), 1082.

(34) Dobrov, S. V. On the Spin-Fermion Connection. *J. Phys. A. Math. Gen.* **2003**, *36* (39), L503.

(35) Ring, P.; Schuck, P. *The Nuclear Many-Body Problem*; Springer Science & Business Media, 2004.

(36) Chen, G. P.; Scuseria, G. E. Robust Formulation of Wick's Theorem for Computing Matrix Elements between Hartree–Fock–Bogoliubov Wavefunctions. *J. Chem. Phys.* **2023**, *158* (23).

(37) Lin, H. Q. Exact Diagonalization of Quantum-Spin Models. *Phys. Rev. B* **1990**, *42* (10), 6561.

(38) Ummethum, J. Calculation of Static and Dynamical Properties of Giant Magnetic Molecules Using DMRG. Dissertation, Universität Bielefeld, 2012.

(39) Mayer, I. The Spin-Projected Extended Hartree-Fock Method. In *Advances in quantum chemistry*; Elsevier, 1980; Vol. 12, pp 189–262.

(40) Majumdar, C. K.; Ghosh, D. K. On Next-Nearest-Neighbor Interaction in Linear Chain. I. *J. Math. Phys.* **1969**, *10* (8), 1388–1398.

(41) Caux, J.-S. Correlation Functions of Integrable Models: A Description of the ABACUS Algorithm. *J. Math. Phys.* **2009**, *50* (9).

(42) Rousochatzakis, I.; Laeuchli, A. M.; Mila, F. Highly Frustrated Magnetic Clusters: The Kagomé on a Sphere. *Phys. Rev. B* **2008**, *77* (9), 94420.

(43) Jiménez-Hoyos, C. Variational Approaches to the Molecular Electronic Structure Problem Based on Symmetry-Projected Hartree–Fock Configurations, PhD Thesis, Rice University, 2013.

(44) Löwdin, P.-O. Quantum Theory of Many-Particle Systems. III. Extension of the Hartree–Fock Scheme to Include Degenerate Systems and Correlation Effects. *Phys. Rev.* **1955**, *97* (6), 1509.

(45) Delfs, C.; Gatteschi, D.; Pardi, L.; Sessoli, R.; Wieghardt, K.; Hanke, D. Magnetic Properties of an Octanuclear Iron (III) Cation. *Inorg. Chem.* **1993**, *32* (14), 3099–3103.

(46) Waldmann, O. Symmetry and Energy Spectrum of High-Nuclearity Spin Clusters. *Phys. Rev. B* **2000**, *61* (9), 6138.

(47) Schnalle, R.; Schnack, J. Calculating the Energy Spectra of Magnetic Molecules: Application of Real-and Spin-Space Symmetries. *Int. Rev. Phys. Chem.* **2010**, *29* (3), 403–452.

(48) Ghassemi Tabrizi, S.; Kühne, T. D. Simultaneous Spin and Point-Group Adaptation in Exact Diagonalization of Spin Clusters. *Magnetism* **2025**, *5* (1), 8.

(49) Klemm, R. A.; Efremov, D. V. Single-Ion and Exchange Anisotropy Effects and Multiferroic Behavior in High-Symmetry Tetramer Single-Molecule Magnets. *Phys. Rev. B* **2008**, *77* (18), 184410.

(50) Ghassemi Tabrizi, S. Symmetry-Induced Universal Momentum-Transfer Dependencies for Inelastic Neutron Scattering on Anisotropic Spin Clusters. *Phys. Rev. B* **2021**, *104* (1), 14416.